\documentclass{article}\usepackage[]{graphicx}\usepackage[]{color}
\makeatletter
\def\maxwidth{ %
  \ifdim\Gin@nat@width>\linewidth
    \linewidth
  \else
    \Gin@nat@width
  \fi
}
\makeatother

\definecolor{fgcolor}{rgb}{0.345, 0.345, 0.345}

\usepackage{framed}
\makeatletter
\newenvironment{kframe}{%
 \def\at@end@of@kframe{}%
 \ifinner\ifhmode%
  \def\at@end@of@kframe{\end{minipage}}%
  \begin{minipage}{\columnwidth}%
 \fi\fi%
 \def\FrameCommand##1{\hskip\@totalleftmargin \hskip-\fboxsep
 \colorbox{shadecolor}{##1}\hskip-\fboxsep
     \hskip-\linewidth \hskip-\@totalleftmargin \hskip\columnwidth}%
 \MakeFramed {\advance\hsize-\width
   \@totalleftmargin\z@ \linewidth\hsize
   \@setminipage}}%
 {\par\unskip\endMakeFramed%
 \at@end@of@kframe}
\makeatother

\definecolor{shadecolor}{rgb}{.97, .97, .97}
\definecolor{messagecolor}{rgb}{0, 0, 0}
\definecolor{warningcolor}{rgb}{1, 0, 1}
\definecolor{errorcolor}{rgb}{1, 0, 0}
\newenvironment{knitrout}{}{} 

\usepackage{alltt}
\usepackage[stable]{footmisc}
\usepackage{fullpage}
\usepackage{setspace}
\usepackage[]{graphicx}
\usepackage[]{color}
\usepackage[OT1]{fontenc}
\usepackage{amsthm,amsmath,amsfonts}
\usepackage{natbib}
\usepackage[colorlinks,citecolor=blue,urlcolor=blue]{hyperref}
\usepackage{bm}
\usepackage{subcaption}
\usepackage{graphicx}
\usepackage{multirow}
\usepackage{xr}
\usepackage{tikz}
\usepackage{authblk}
\usepackage[section]{placeins}

\newcommand{\EE}{\mathbb{E}}

\newcommand\numberthis{\addtocounter{equation}{1}\tag{\theequation}}

\newenvironment{ass}[2][Assumption:]{\begin{trivlist}
\item[\hskip \labelsep {\bfseries #1}\hskip \labelsep {\bfseries #2}.]}{\end{trivlist}}

\def\independenT#1#2{\mathrel{\rlap{$#1#2$}\mkern2mu{#1#2}}}
\newcommand\independent{\protect\mathpalette{\protect\independenT}{\perp}}

\newcommand{\indicator}[1]{\mathbf{1}_{\left[ {#1} \right] }}

\renewenvironment{knitrout}{\begin{singlespace}}{\end{singlespace}}

\title{Student Log-Data from a Randomized Evaluation of Educational
  Technology: A Causal Case Study}

\author[1]{Adam C Sales}
\author[2]{John F Pane}
\affil[1]{University of Texas, Austin, TX}
\affil[2]{RAND Corporation, Pittsburgh, PA}

\IfFileExists{upquote.sty}{\usepackage{upquote}}{}
\begin{document}
\maketitle

\section{Introduction}
From 2007 through 2010, the RAND corporation
conducted one of the first effectiveness trials awarded through
competitive grant programs sponsored by the US Department of
Education's Institute of Education Sciences.  The randomized
controlled trial (RCT) was
designed to estimate the effect of school-wide adoption of the
Cognitive Tutor Algebra I (CTA1) curriculum, whose centerpiece is a
computerized ``tutor'' that uses cognitive science principles to teach
Algebra I \citep{anderson1985intelligent}.
The study \citep{pane2014effectiveness} found no effects in its first year, but in
the second year of implementation students in high schools randomized to
the CTA1 condition outperformed the control group on the posttest by a fifth of a
standard deviation.
As educational technology (EdTech) booms, so do RCTs in the mold of the RAND
CTA1 study. A recent systematic review \citep{escueta2017education} cites 29
published reports of RCTs of ``computer assisted learning'' programs,
all but one of which was published since 2001. We are aware of
numerous other such RCTs planned or ongoing.

RCTs are invaluable tools for estimating these programs'
efficacy. However, the interventions they study are highly
multifaceted and complex. Each classroom, and each student within
those classrooms, may use an EdTech application in a different way;
presumably, the application's efficacy depends on these usage
decisions.
Similarly, EdTech applications often include a number of optional
features---to what extent do each of these drive the program's
effectiveness?
The data collected during typical education RCTs---data on treatment
assignments, outcomes, a standard set of demographics, and possibly a
pretest---offer limited options for studying treatment effect
heterogeneity, and offer no information on how the intervention was
implemented.

However, with little extra effort, researchers studying EdTech can
gather rich implementation data over the course of an RCT.
For instance, the software administrators running the CTA1 RCT
gathered computer log data for students in the treatment group.
They assembled a dataset that recorded which problems each student
worked, along with timestamps, the numbers of hints requested, and the
number of errors committed for each problem.
Analogous log data is (or may be) collected from other EdTech RCTs.
As one example, RAND recently
completed an efficacy study of a different algebra tutoring system,
ALEKS, and gathered student log data in order to study
implementation.

Analysis of log data from EdTech RCTs could, in principle, lead to
insights on the relationship between implementation and effectiveness.
Using log data, researchers could investigate how the product was
actually implemented, as well as what aspects of implementation
correlate with higher (or lower) treatment effects, or which features
of the program drive its effects.
For instance, \citet{sales2016student,aoas,reassignmentEffect}
investigated aspects of mastery learning with CTA1.
An ongoing research project uses the ALEKS log data to
study its instructional material: when students begin working on a particular topic in ALEKS, the
system first presents instructional material, followed by problem
solving activities, and then, if students are unsuccessful at solving
the problems, more instructional material. Some students diligently
read the initial instructional material while others skip it and dive
directly into problem solving. Is skipping the material a productive
strategy for students in terms of learning algebra, or should ALEKS be
redesigned to prevent such skipping?

Answering questions like these calls for new statistical tools, which must simultaneously surmount three challenges: first,
implementation is not typically randomized, so the relationship
between log data and outcomes is likely confounded. Second,
log data is only gathered from students assigned to use the
intervention, and not from the control group. Third, log datasets have
a complex structure---often a large number of granular observations of several types
gathered over the course of each student's engagement with the
program.

This paper will use log data from the CTA1 study---specifically, data on students' use
of hints within the software---to illustrate three approaches to these
challenges.
We will attempt to answer two related questions: first, what is the role of
the availability of hints in driving the overall CTA1 treatment
effect? Second, can requesting hints more often lead to a larger
treatment effect?
The first approach we consider discards the control group, and analyzes
hint-requesting in the treatment group as an observational study,
using a matching design
\citep[c.f.][]{rosenbaum2002observational}. The
second approach applies the framework of causal mediation analysis
\citep{vanderweele2015explanation,hong2015causality,imai2011unpacking} to estimate the role of
hint-requesting in the overall CTA1 treatment effect. However, since
students in the control group were unable to request hints, not all
mediational estimands are identified, and estimation methods must be
modified.
Finally, we conduct a principal stratification analysis
\citep{frangakis} to learn if students requesting hints at different
rates experienced different treatment effects.

Our intention is to develop, demonstrate, and contrast these three
approaches to modeling log data.
A rich, and often polemic, literature surrounds principal
stratification and mediation analysis
\citep[e.g.][]{rubin2004direct,vanderweele2011principal,pearl2011principal,mealli2012refreshing,vanderweele2012comments}.
\citet{vanderweele2008simple}, \citet{jo2008causal} and others have
given conditions under which certain principal stratification and mediational
estimands coincide.
Our contribution will build on this literature in two ways: first, by
providing a detailed demonstration and comparison of matching, mediation analysis,
and principal stratification on the same dataset, and second by
focusing on the applicability of these three methods in the specific
context of studying implementation data from an randomized
experiment studying educational technology.

The following section will provide background for the case-study,
describing the CTA1 curriculum, the RAND study, and our focus on
hints.
The next three sections will present the observational study,
causal mediation, and principal stratification, respectively.
As it turns out, these analyses appear to give contradictory
results---the observational study and mediation analysis estimate a
negative effect of requesting hints, while the principal
stratification analysis finds that students who requested more hints
may have experienced larger overall treatment effects.
Section \ref{sec:synthesis} will contrast these findings (including a
possible reconciliation), compare the three methods generally,
and discuss how they may be applied to data from other educational
technology RCTs.
Section \ref{sec:conclusion} will conclude.

The \texttt{R} and Stan code used to for all of the analyses in the paper can be found in a github repository, \url{https://github.com/adamSales/logDataCaseStudy}.

\section{Case Study: The Role of Hints in the Cognitive Tutor Effect}

\subsection{The Cognitive Tutor}
CTA1 is one of a series of complete mathematics curricula developed by
Carnegie Learning, Inc., which include both textbook materials and an
automated computer-based Cognitive Tutor
\citep{anderson1995cognitive,pane2014effectiveness}.

The computerized tutor was originally built to test the ACT and ACT-R
theories of cognition \citep{anderson2013architecture}.
These are elaborate theories that describe, among other things, the necessary components of
cognitive skills in, say, mathematics or computer programming, and the
process of acquiring those skills.
The Algebra I tutor guides students through a sequence of Algebra I
problems nested in sections within units, and organized by the specific sets of skills they require.
Students move through the curriculum as the tutor's internal learning
model determines that they have mastered the requisite skills.

\subsection{The Role of Hints\footnote{This subsection draws heavily
  on comments from an anonymous reviewer.}}
Students using the Cognitive Tutor Algebra software work algebra
problems on a computer.
Each problem in the software is associated with a set of skills that
the student must master in order to solve the problem.
The software also specifies a ``solution path'' for each problem---a sequence
of actions (such as ``multiply both sides of the equation by 3'')
students must take in order to solve the problem.
If there is more than one way to solve a problem, the software specifies
multiple correct solution paths students may follow.
As soon as a student errs in working through a problem---that is, departs from one of the
solution paths---the software shows the student an error message.
This feedback is tailored to the specific mistake, and
designed to guide the student back to a correct solution path.
Similarly, when a student gets stuck, and is unable to determine the
next step in solving the problem, he or she can ask for a hint.
Like the error feedback, these hints are tailored to the specific
algebra skills corresponding to the next step on the solution path.
Thus, hints and error feedback are crucial aspects of the tutor's
pedagogy.

There are good reasons to believe that the availability of
hints---help on demand---may increase student learning, but there is
also reason to be skeptical \cite[See][for an overview of the theory]{aleven2016help}.
First of all, as, e.g., \citet{anderson1995cognitive} points out, ``if
students never receive help of any sort, they are in danger of becoming
permanently stuck on some problem'' (p. 190).
That is, hints give students who are stuck a way to continue working.
Hints may also boost learning by ``helping students identify relevant features in
problems'' \citep[][p. 6]{aleven2016help},  that is, point students
towards aspects of the problem that are ``most important for the goals
of learning'' \citep[][p. 782]{koedinger2012knowledge}.
Hints can help improve students' self-awareness and
problem-solving strategy, such as by clarifying which skills they have
and have not mastered.

Most problems are associated with several hints, arranged in a
sequence so that a student who remains stuck after one hint may
request another.
The last hint in the sequence will show a complete solution to the
problem---essentially showing students a worked example, which can be an
effective tool for learning \citep[e.g.][]{sweller1985use}.

On the other hand, all of these theoretical mechanisms rely on
students to play their part.
In order for hints to be instructive, students need to reflect on
their content in productive ways.
This suggests that hints may be helpful for some students but not for
others.
In the same vein, hints need to be well-designed in order to be
effective \citep{mckendree1990effective}, so some hints may be more
helpful than others.

Hints may even be harmful: as \citet{koedinger2007exploring} put it
(p. 241),
``many lines of research and theory suggest the importance of \dots
withholding information from students so that they can exercise, test,
or reason toward new knowledge on their own.''
\citet{paas1994variability} interpret this dilemma in terms of balancing
``cognitive load,'' that is, optimally allocating a student's
attention to the most relevant aspects of a problem.

The empirical literature on the value of hints in intelligent tutors
is mixed \citep[see][for a summary]{goldin2012learner}.
\citet{anderson1995cognitive} compared the performance of students
randomly assigned to different versions of the Cognitive Tutor with
different feedback structures, and found that students who were able
to request hints finished the lesson faster than those who were not,
but could not detect an effect on learning.
\citet{singh2011feedback} used a similar design, comparing two
versions of the ASSISTments intelligent tutor, one in which hints and
immediate error messages were available, and one which did not provide
feedback. Students in the feedback condition showed larger gains on a
posttest, though it is unclear whether these gains were due to hints
or error feedback.
\citet{beck2008does} and \citet{goldin2012learner}, did not randomize hint
availability, but instead used statistical models to control for
observed confounding.
Both papers used students' subsequent performance within the tutor to
estimate hint effects.
\citet{beck2008does} found conflicting results using different
analytical methods to control for confounding.
\citet{goldin2012learner} investigated two sources of heterogeneity in
the effect of a hint request: by student and by the ``level'' of the
hint (the first, second, or final hint available for a given problem).
They found that the effect of asking for a hint varies
between students, but on average it is negative for the first two
hints requested on a problem, but positive for the final hint.

\subsection{Measuring Hint Usage in the CTA1 Data}
Since the purpose of the CTA1 study was to estimate the overall effect
of the CTA1 curriculum---including the associated textbook and
recommended classroom practices---access to the full CTA1 curriculum
was randomized, rather than any specific aspect of usage.
Effectiveness was measured with a
standardized posttest covering a broad range of Algebra I skills.
In this regard, the CTA1 study is similar to other large-scale
evaluations of educational technology which do not randomize usage but
do include a posttest.

The CTA1 study was of a much longer duration than
any of the studies of hint effects we are aware of: students had access to the tutor for an entire
school year before the posttest.
This factor allows for much more data for each student than shorter
studies, as well as more variability in the types of problems and
contexts in which students may request hints.

In order to align with our research focus, we narrow the type
of hint data we model.
Since our posttest measured Algebra I skills, we only
considered worked problems from the Algebra I curriculum.
Next, although \citet{goldin2012learner} demonstrated the importance
of hint level, our focus is on the overall average effect of
requesting any hints; therefore, the variable measuring a student's
hint request on a problem was set to one if the student requested any
hints on that problem, and zero otherwise.

Lastly, any observational study of hint usage must contend with the
fact that students are unlikely to request hints on problems that they
already know how to solve.
Hence, any measure of hint usage is necessarily also a measure of
student ability---students who know more algebra will almost certainly
request fewer hints.
This leads to two related problems: first, student ability
confounds any statistical relationship between hint request frequency
and posttest scores.
Secondly,
a student's decision to request a hint implies both that she
found the problem at least somewhat difficult, and that, in this case,
she responded to that difficulty by requesting a hint.
Our interest is in the second implication, not the first.
Conversely, the significance of a student's decision to forgo a hint
depends in part on how challenging he or she finds the problem.
Hint request frequency combines students' hint requests (or non-requests) from both
problems they find challenging and trivial, so is a poor measure of
students' disposition to use the hint feature.

Ideally, we would solve this problem by only including data on
problem-student pairs in which the problem was challenging to the
student.
Of course, no direct measure of problem challenge is available.
Instead, we considered a problem to be challenging if the student
either requested a hint or made an error on that problem.
Then, let
\begin{equation*}
\bar{h}_i=\frac{\text{\# problems on which }i\text{ requested a
    hint}}{\text{\# problems on which }i\text{ either requested a hint
    or made an error}}
\end{equation*}
measure hint frequency.

This measure is inversely related to the ratio of the percentage (of
all
problems) a student makes an error without having requested a hint,
to the percentage of problems a student requests a hint.%
\footnote{Technically, if $h$ is the event that a student requests
a hint on a problem and $e$ is the event that the student makes
an error, then $Pr(h|h\text{ or }e)=1/\{1+Pr(e\text{ and not }h)/Pr(h)\}$.}
Making an error on a problem without having requested a hint reflects
a hesitance to request a hint when one might have been warranted (or a
careless mistake).
$\bar{h}$ ignores the class of challenging problems on which students
are challenged and figure the answer out (or guess correctly) without
requesting a hint.
This class of problems is inherently interesting to our question;
however, it is unidentified.

\subsubsection{Dichotomizing Hint Usage}\label{sec:dichotomizingHintUsage}
The observational study and mediation analysis below require a dichotomous measurement of hint usage: is a
student a high or low hint user?
To this end, we dichotomize $\bar{h}$ by comparing it to a cutoff value, so
that $H_i=\indicator{\bar{h}_i>c}$, where $\indicator{x}=1$ if $x$ is true
and 0 otherwise.

To choose $c$, we first fit a modified Rasch mixture model
\citep{rasch} to hint request data.
Specifically, we modeled the probability of at least one hint request in each problem
$p$, worked by student $i$, in which $i$ requested a hint and/or
committed an error, as
\begin{equation}\label{eq:rasch}
Pr(h_{ip})=logit^{-1}(\eta_i+\delta_{s[p]})
\end{equation}
where $h_{ip}$ is the event that student $i$ requests at least one hint on
problem $p$, $\eta_i$ is a student parameter, $\delta_{s[p]}$ is a
section-level parameter for $s[p]$, the section that problem $p$ was
drawn from, and $logit^{-1}(x)=\{1+exp(-x)\}^{-1}$ is the inverse logit
function.

Typically, the student parameter $\eta_i$ measures student ability;
here, instead, it measures a student's proclivity to request a hint.
The typical Rasch model includes a problem-level ``difficulty''
parameter; here, it would measure the likelihood an individual problem
elicits a hint request.
That parameter is important in our context because the
specific problems a student works on may influence his or her hint
requests.
For instance, two students with the same underlying proclivity to request
hints $\eta$ may actually differ in their observed hint requests, because one
worked on harder problems than the other.
However, in our dataset many problems were worked by very few students, so problem-level parameters would be hard to estimate.
Instead, we included the section-level parameter $\delta_{s[p]}$,
which measures the extent to which students tend to request hints on
problems in section $s$.
Ignoring the differences between problems within a section could allow for wider
variance in the average problem difficulty each student experiences
than is captured by the section level parameters $\delta_{s[p]}$.
That could, in turn, induce a difference in hint requests between two
students with similar $\eta$ or vice-versa, and lead
to bias in our estimate of $\eta$.
Since the Cognitive Tutor selects problems for students based on estimates
of their current skill mastery \citep{steveFaied2020}, this
possibility is not out of the question.

With even more granular data, \eqref{eq:rasch} could be further
elaborated, avoiding this assumption.
In the Cognitive Tutor, problems often have multiple parts, and each part offers its own hints.
Researchers in possession of data on each part of each worked problem could specify a model at the part-level, instead of at the problem-level like \eqref{eq:rasch}.
Perhaps an even better option (suggested by an anonymous reviewer) would incorporate data on the specific skills required to solve each problem.
While sections of the tutor differ from each other in the skills they teach, not every problem in each section includes the same set of skills; different problems requiring the same skill set will be similar in difficulty \citep[see, e.g.]{goldin2012learner,pavlik2009performance}.

In order to dichotomize hint request behavior, we modeled
student effects $\eta$ as
\begin{equation}\label{eq:mixture}
\eta\sim
p_0\mathcal{N}(\mu_0,\sigma_0)+(1-p_0)\mathcal{N}(\mu_1,\sigma_1)
\end{equation}
and constrained $\mu_0<\mu_1$.
This mixture model clusters students into two categories: high hint
users have $\eta$ drawn from a normal distribution with mean $\mu_1$
and standard deviation $\sigma_1$, and low hint users have $\eta$
drawn from a normal distribution with mean $\mu_0$ and standard
deviation $\sigma_0$.

Our main parameter of interest here is $p_0$, the proportion of
students who are low hint users.
We estimated $p_0\approx$0.7,
classifying 30\% of students as
high hint users, so we chose $c$ as the
70th percentile of $\bar{h}$,
approximately 0.6.
This definition agreed with the model's classification (based on
$\eta$ as opposed to $\bar{h}$) in
approximately 93\% of cases.

The Rasch model for hint usage \eqref{eq:rasch}, but not the mixture
model \eqref{eq:mixture} will play an important role in the principal
stratification analysis of Section \ref{sec:principalStratification}.

\subsection{Data}\label{sec:data}

\citep{pane2014effectiveness} estimated effects separately in middle
schools and high schools in each year of implementation; the
only statistically significant treatment effect was in the high school
sample in the second year.
Since our goal was to better understand the CTA1
treatment effect, we focused our analysis on data from high school students in
the second year of the CTA1 trial, for whom the treatment effect was
most evident.

We merged data from two sources: computerized log data gathered by
Carnegie Learning, and covariate, treatment and outcome data gathered by RAND.

The log data lists the problem name, section, and unit of
each problem, the numbers of hints requested and errors committed, and time-stamps.
Log data were missing for some students, either because the log files
were not retrievable, or because of an imperfect ability to link log
data to other student records.
Further, log data for sections that were not part of the
standard CTA1 Algebra I curriculum and sections worked by fewer than 100
students were omitted from the
dataset.\footnote{The principal stratification model was re-run without dropping sections, and
  after dropping sections worked by fewer than 500 students, with
  similar results.}

To construct our analysis dataset, we first dropped treatment schools with log data missing
for 90\% or more students.
Prior to randomization, schools were stratified into pairs or triples
based on baseline covariates, and randomized between treatment and
control condition within those randomization blocks.
When we dropped a treatment school from our analysis, we also dropped
the control schools in its randomization block.
Of the remaining 2,390 students, 88\% (2,108) had log
data.
Then, for the sake of simplicity, the treatment students without log data
were dropped from the study.\footnote{Including these students in a
  principal stratification model is straightforward \citep{aoas}. Including
  subjects with missing log data in a mediation or observational study
  design can be more problematic \citep[see, e.g.][]{li2017identifiability}.}

All told, the analyses presented here were all based on 2,108 students assigned to the CTA1 condition; the estimation of direct effects in Section \ref{sec:mediation} and the principal stratification also relied on the 2,918 students assigned to control.
Together, the 5,026 students were nested within 116 teachers, in 43 schools across five states.

Table \ref{tab:covariateBalance} describes the covariates we used, including
missingness information, control and treatment means, and standardized
differences \citep[c.f.][]{kalton1968standardization} from the final
analysis sample.
We singly-imputed missing values with the Random Forest routine implemented
by the \texttt{missForest} package in \texttt{R}
\citep{missForest,rcite},  which estimated ``out of box'' imputation
error rates as part of the random forest regression, also shown in Table \ref{tab:covariateBalance}.

\begin{table}[ht]
\centering

\begin{tabular}{cccrccc}
  \hline
  \hline
 &\% Miss.& Imp. Err.&Levels& BaU & CTA1 & Std. Diff.\\
  \hline
  \hline
\multirow{2}{*}{Grade}&\multirow{2}{*}{3\%} &\multirow{2}{*}{0.01}&9th&91\% &90\% &-0.04\\
&&&$>$9th&9\% &10\% &0.04\\
\hline
\multirow{2}{*}{ELL}&\multirow{2}{*}{4\%} &\multirow{2}{*}{0.01}&No&95\% &96\% &0.09\\
&&&Yes&5\% &4\% &-0.09\\
\hline
\multirow{2}{*}{FRL}&\multirow{2}{*}{29\%} &\multirow{2}{*}{0.29}&No&26\% &27\% &0.02\\
&&&Yes&74\% &73\% &-0.02\\
\hline
\multirow{3}{*}{Ethnicity}&\multirow{3}{*}{6\%} &\multirow{3}{*}{0.23}&White/Asian&47\% &52\% &0.16\\
&&&Black/Multi&32\% &26\% &-0.14\\
&&&Hispanic/Nat.Am.&21\% &22\% &-0.03\\
\hline
\multirow{2}{*}{Sex}&\multirow{2}{*}{5\%} &\multirow{2}{*}{0.35}&Female&51\% &49\% &-0.04\\
&&&Male&49\% &51\% &0.04\\
\hline
\multirow{3}{*}{Sp. Ed.}&\multirow{3}{*}{1\%} &\multirow{3}{*}{0.11}&Typical&87\% &86\% &-0.00\\
&&&Sp. Ed.&8\% &8\% &-0.02\\
&&&Gifted&5\% &6\% &0.03\\
\hline
Pretest&16\%&0.20& &-0.33& -0.36& -0.05\\
\hline
&&&\multicolumn{4}{c}{Overall Covariate Balance: p=0.55}\\
\hline
\hline
\end{tabular}

\caption{Missingness information, control (``BaU'' or ``Business as
  Usual'') and treatment (``CTA1'') means, and balance for the
  covariates included in this study, from the high school year two
  stratum of CTA1 Effectiveness experiment. Imputation error is percent falsely classified for
  categorical variables (Race/Ethnicity, Sex, and Special Education)
  and standardized root mean squared error for Pretest, which is
  continuous. 
  Analysis done in \texttt{R} via \texttt{RItools} \citep{ritools}.}
\label{tab:covariateBalance}
\end{table}

\section{An Observational Study Within an Experiment}\label{sec:observational}

How does requesting hints within CTA1 affect learning?
More precisely, would low hint users have achieved higher posttest scores
had they requested hints more often?
Would high hint users have achieved higher posttest scores had the
requested hints less often?

This section will illustrate an observational study to answer these
questions.
Generally speaking, the observational study approach discards the
control group entirely and uses statistical confounder control (in our
case, propensity score matching) to estimate causal effects of a usage parameter on posttest scores.
Implementing the method requires data from members of the treatment group: usage (in our case, $H$), posttest scores or another outcome of interest, and a set of baseline covariates to control for confounding.
Since the control group is not used, identification does not depend on randomization---in fact, this same approach could be (and is) used with non-experimental data.
However, outside of a planned experiment, posttest scores may not be available.
Our usage variable $H$ is binary, which is necessary for classical propensity score matching; alternative methods are available for usage variable that are continuous \citep[e.g.][]{hirano2004propensity}, ordered \citep[e.g.][]{leon2005mixed}, categorical \citep[e.g.][]{lopez2017estimation} or other types.
Like all observational studies, it depends on the strong, untestable assumption of no unmeasured confounding.

The approach we take here, following \citet{rosenbaum2002observational},
 \citet{hansen2009b}, and \citet{ho:etal:2007}, has three broad steps.
In the first, we construct a ``match," identifying groups of students, some of whom have $H=0$ and others with $H=1$, but who are otherwise comparable.
The hope is that the distribution of $H$ conditional on this match
resembles what might have been seen if (counterfactually) $H$ had been
randomized within matched groups.
Thus we evaluate the success of the match by checking if matched
students have similar baseline covariate distributions; if necessary,
we may revise match.
In the second step, we estimate the average effect of $H$ on posttest scores, by comparing posttest scores between matched students with $H=1$ and $H=0$.
In the third and final step, we account for the possibility of unmeasured confounding in a sensitivity analysis.

The next subsection will give more formal background on propensity score matching, including notation and identification assumptions. Following subsections will illustrate each of the three steps of propensity score matching.

\subsection{Observational Study: Background}
Let $Y_i$ represent subject $i$'s posttest score, and let $Z_i$ represent
$i$'s treatment status (i.e. the CTA1 or control group).
Following \citet{neyman} and \citet{rubin}, let $Y_i(Z=1)$ be
the posttest score student $i$ would achieve were $i$ (perhaps
counterfactually) assigned to the treatment condition, and
$Y_i(Z=0)$ the score were $i$ assigned to control.
Then the CTA1 treatment effect for student $i$ is $Y_i(Z=1)-Y_i(Z=0)$.
For students in the treatment group, define potential outcomes $Y_i(Z=1,H=1)$ and
$Y_i(Z=1,H=0)$, or $Y_i(H=1)$ and $Y_i(H=0)$ for short, corresponding
to $i$'s posttest score were $H_i=1$ or 0, respectively.
Then the effect of $H$ on $Y$ for student $i$ is $Y_i(H=1)-Y_i(H=0)$.
This structure implicitly assumes ``non-interference,'' that one
student's treatment assignment $Z$ or hint proclivity $H$ does not
affect another student's posttest scores.

Individual treatment effects are (typically) unidentified, since only
one of the relevant potential outcomes is ever observed for each
subject.
For instance, $Y(H=1)$ is unobserved for members of the
treatment group with $H=0$.
If $Pr(H_i=1)$ were known for each $i$, and bounded away from 1 and 0, (as would be the case if $H$
were randomized) then the average treatment effect (ATE) of $H$ on $Y$,
$\EE[Y(H=1)]-\EE[Y(H=0)]$, could be estimated without bias or further
assumptions.
Of course, the distribution of $H_i$ is unknown, and is presumably a
function of $i$'s individual characteristics (which, as we shall see,
differed at baseline between those students with $H=1$ and those with $H=0$).

Let $\bm{x}_i$ represent a vector of baseline covariates for subject $i$.
In our study of hints, $\bm{x}$ includes indicators for state, school,
and classroom, and the variables described in Table \ref{tab:covariateBalance}.
Then we assume \citep[c.f.][]{rosenbaum1983central}:
\begin{ass}{Strong Ignorability}
\begin{equation*}
 \{Y(H=1),Y(H=0)\}\independent H |\bm{x}
\end{equation*}
\end{ass}
that conditional on $\bm{x}$, potential outcomes $Y(H=1)$ and $Y(H=0)$
are independent of realized $H$.
Under strong ignorability, one may compare subjects with identical
covariates $\bm{x}$ to estimate causal effects.

Unfortunately, this sort of exact matching is impossible in our finite
sample.
Instead, we estimated ``propensity scores'' \citep{rosenbaum1983central}:
\begin{equation*}
\pi_i=Pr(H_i=1|\bm{x})
\end{equation*}
the probability of a treatment-group subject requesting frequent hints, conditional on
his or her covariate vector $\bm{x}$.
\citet{rosenbaum1983central} shows that conditioning on $\pi$ is
equivalent to conditioning on $\bm{x}$.
We estimated $\pi$ using a multilevel logistic regression,
using the \texttt{lme4} package in
\texttt{R} \citep{lme4}.
The propensity score model was:
\begin{equation*}
\pi_i=logit^{-1}(\alpha+\bm{\tilde{x}}\bm{\beta}+\epsilon^{state}+\epsilon^{school}+\epsilon^{class})
\end{equation*}
where $\bm{\tilde{x}}$ is a vector of covariates including the
variables in Table \ref{tab:covariateBalance}, missingness
indicators for grade, race, sex, and economic disadvantage, and
a natural spline with five degrees of freedom for pretest.
State, school, and class were each included using normally
distributed random intercepts, $\epsilon^{state}$,
$\epsilon^{school}$, and $\epsilon^{class}$.

\subsection{Constructing and Evaluating a Match}

We constructed an optimal full matching design \citep{hansen2004} to
condition on estimated propensity scores $\hat{\pi}$ using the
\texttt{optmatch} package in \texttt{R} \citep{optmatch}.
We matched $H=0$ subjects to $H=1$ subjects in such a way as
to minimize the overall distance between subjects in matched sets in
the logit-transformed propensity scores, $logit(\hat{\pi})$.
We constrained the match so that students could only be matched within
schools, since schools determine a number of factors
that in turn may impact posttest scores, including baseline
student characteristics, pedagogical styles, and CTA1 usage patterns
\citep[see][]{descriptivePaper}. 
Each matched set was allowed to contain any positive number of $H=1$
and any positive number of $H=0$ subjects, resulting in matches of
variable size and composition.
For instance, one matched set included 44 $H=1$ subjects matched to a
single $H=0$ subject, and another matched 117 $H=0$ subjects to a
single $H=1$ subject.

This strategy allowed every student in the CTA1 arm our dataset to be included in the match.
In contrast, many matching studies discard subjects in the data sample
with estimated propensity scores close to 1 or 0, focusing on the
``region of common support'' \citep[e.g.][]{caliendo2008some,shadish2010primer}.
Doing so requires modifying the causal estimand---studies can only
estimate the ATE for included subjects and those who resemble them.
We chose, instead, to include every subject in order to simplify the mediation
analysis in the following section, and validated the match by
inspecting covariate balance.
This came at the cost of imprecision in estimating the overall ATE,
which we discuss in more detail below.

Figure \ref{fig:balance} and Table \ref{tab:balance} in the online appendix show that
matching largely eliminated mean differences in pre-treatment
covariates between $H=1$ and $H=0$ students.
The only notable exception is in pretest scores: $H=1$ students tended
to have slightly higher pretest scores than their matched comparisons
with $H=0$.
The p-value from an omnibus covariate balance test
\citep{covBal} is 0.39.

\subsection{Estimating the Effects of $H$ on Posttest Scores}\label{sec:obsEst}

Let $M_i$ be a categorical variable denoting the matched set that $i$
belongs to.
Then we may re-state the ignorability condition with reference to this
match \citep[c.f.][]{rebar}:
\begin{ass}{Matched Ignorability}
\begin{equation*}
 \{Y(H=1),Y(H=0)\}\independent H |M
\end{equation*}
\end{ass}
Under matched ignorability, the difference of mean outcomes between $H=1$ and
$H=0$ students within each match is unbiased for the average treatment
effect for students in that match.
In other words, if $\tau_m=\overline{Y(H=1)}-\overline{Y(H=0)}$ is the
average effect in match $m$, then
$\hat{\tau}_m=\bar{Y}_{M=m,H=1}-\bar{Y}_{M=m,H=0}$ is an unbiased
estimate of $\tau_m$, where $\bar{Y}_{M=m,H=h}$ is the average
observed outcome for subjects in match $m$ with hint level
$h\in\{0,1\}$.

Weighted means of those estimated treatment effects, of the form
\begin{equation}\label{eq:tauWeightedMean}
\hat{\tau}_w=\sum_m w_m\hat{\tau}_m/\sum_m w_m
\end{equation}
estimate aggregate
treatment effects in the sample.
To estimate the average effect of $H$ for all subjects, set $w_m=
n_{1m}+n_{0m}$, where $n_{1m}$ and $n_{0m}$ are the number of
$H=1$ and $H=0$ subjects in match $m$, respectively.
Alternatively, setting $w_m=n_{1m}$ estimates the the ``treatment on the
treated,'' or TOT, effect---the average effect of
$H$ on those subjects for whom $H=1$.
Weights $w_m=(1/n_{1m}+1/n_{0m})^{-1}$ are ``OLS'' or
``precision'' \citep[c.f.][]{schochet2015statistical}; the
estimate using these weights is equal to the coefficient on $H$ from
an ordinary least squares (OLS) regression of $Y$ on $H$ and a set of dummy
variables for $M$.
Under standard OLS assumptions, precision weights minimize the
standard error of the weighted mean estimator.


Table \ref{tab:matchResults} gives treatment effect estimates,
standard errors, and confidence intervals under these three weighting
schemes.
All three sets of standard errors
and confidence intervals used the heteroskedasticity-robust ``HC3'' sandwich estimator
\citep{sandwichPackage}.
The three estimates all roughly agree that being a high hint user
decreases posttest scores by around 0.15 standard deviations.
That is, they suggest that requesting hints hurts posttest scores.

The similarity of the three estimates boosts our confidence, if only
slightly. The ATE and TOT estimates refer to different groups
of students---ATE estimates an average effect for the entire treatment
group, whereas TOT estimates an effect for high hint users. Had these
been very different, one might wonder why the effect of hint requests
varies so much between high and low hint users, and if that suggests
selection bias.
The OLS estimate downweights matched sets with many more $H=1$ than
$H=0$ students or vice-versa, relative to matched sets with several of
both.
(Recall that such imbalanced matched sets were the result of the
decision to include every subject in the analysis.)
It is larger in magnitude and much less noisy than the other two
estimates.

\begin{table}[ht]
\centering
\begin{tabular}{rllccc}
  \hline
  &&&&\multicolumn{2}{c}{Sensitivity Intervals}\\
 Weights&Estimate&Std. Error&CI&[Pretest]&[Ethnicity] \\
 \hline
OLS & -0.18 & 0.04 & [-0.25,-0.10] & [-0.41,0.06] & [-0.30,-0.05] \\ 
  ATE & -0.12 & 0.05 & [-0.23,-0.02] & [-0.45,0.20] & [-0.29,0.04] \\ 
  TOT & -0.14 & 0.06 & [-0.25,-0.03] & [-0.47,0.19] & [-0.31,0.03] \\ 
   \hline
\end{tabular}
\caption{Estimates, standard errors, and sensitivity analysis of the weighted average effect of hint usage on posttest scores, under different weighting schemes.} 
\label{tab:matchResults}
\end{table}


\subsection{Sensitivity Analysis: Assessing the Role of Unmeasured Confounding}

None of these estimates addressed the possibility of confounding from
unmeasured covariates; instead, they relied on matched ignorability.
To account for the possibility of unmeasured confounding, we conducted
a sensitivity analysis following the method of \citet{hhh}.
This method imagines a hypothetical missing covariate $U$ and estimates
how the the omission of $U$ alters the estimate
and its standard error.
The result is a ``sensitivity interval'' \citep[c.f.][]{rosenbaum2002observational} that,
with 95\% confidence, contains the true effect accounting for both
sampling uncertainty and uncertainty due to possible confounding.
$U$ is characterized by two sensitivity parameters:
$\rho^2$ is the squared partial correlation between $U$ and $Y$, after
accounting for the observed covariates in the model, and $T_Z$ is the
t-statistic for the coefficient on $U$ from an ordinary least squares
regression of $H$ on $U$ and the observed covariates.
These sensitivity parameters may be benchmarked by estimating their
counterparts when observed covariates are left out of the model.

The last two columns of Table \ref{tab:matchResults} give sensitivity intervals for each of the three weighted average treatment effects.
The column of Table \ref{tab:matchResults} labeled ``[Pretest]''
accounts for the possible omission of a covariate $U$ that, at most, predicts $Y$ and $H$ as well as do pretest scores.
The resulting sensitivity intervals are quite wide, including
both large negative effects as well as substantial positive
effects.
In fact, pretest is the most important of our observed covariates; it is the observed covariate whose omission would cause the most bias.
This is unsurprising, since pretests are generally considered to be the most important covariate to measure, and studies that include pretest measures often reproduce experimental estimates \citep[e.g.][]{cook2008three,cook2009bias}.
Therefore, a sensitivity analysis considering an omitted covariate as important as pretest scores may be too pessimistic.
On the other hand, it is likely that pretest scores do not fully capture between-student variance in academic ability; that is, there may be important components of prior ability (such as the ability to learn algebra) that are correlated with hint requests and posttest scores but not reflected in pretest scores.
Hypothetical measurements of these components could, perhaps,
constitute unobserved covariates with similar or greater importance
than observed pretest scores.

A more optimistic scenario is reflected in the final column of Table \ref{tab:matchResults}.
That column gives sensitivity intervals for the omission of a covariate that, at most, predicts $Y$ and $H$ as well as ethnicity dummy variables, the second most important of our observed covariates.
These sensitivity intervals are also wide, but the interval corresponding
to OLS weights is entirely negative.

The conclusion is that the omission of a confounder as important as ethnicity indicators
cannot explain the estimated OLS-weighted ATE.
Allowing for the possible omission of such a covariate, the results suggest that requesting a
large number of hints hurts students' posttest scores.
The size of the effect may be as small as
5\% of a standard deviation
or as large as 30\% of a standard deviation.

\section{Mediation Analysis}\label{sec:mediation}
The goal of causal mediation analysis is to decompose a treatment
effect into an ``indirect'' or ``mediated''
effect and a ``direct'' effect.
The indirect effect captures the component of the effect that is due
to the mediator: the treatment affected the outcome by affecting the
mediator, which itself, in turn, affected the outcome.
Assignment to the treatment condition allowed students to request many
hints during their work---how did the wide availability of hints affect their posttest scores?
The direct effect is the component of the overall effect that operates via other mechanisms.
For instance, how would assignment to CTA1 affect posttest scores were
hints more limited?

The method we present here builds on the results from the observational study in the previous section.
In fact, in our analysis estimating indirect effects requires no more data than the observational study.
Estimating direct effects requires, additionally, data from the control group: posttest scores and (ideally) baseline covariates.
The randomization of the intervention in the RCT now plays an
important role in identification, since the effects of the
intervention on both hint usage and posttest scores are at issue; the
fact that the control group does not have access to the program
likewise plays a crucial role.
Finaly, the assumption of no unmeasured confounding remains important.

It is worth noting here that conducting mediation analysis requires
measuring or positing a value of the mediator in the control arm of
the study.
We know that CTA1 hints were not available to the control group, so we
set $H=0$ for all members of the control group.
However, other facets of log data may not be coherently defined in
the control group; for instance, students' proclivity to master the
skills of one CTA1 section before moving on to the next \citep{aoas}
or teachers' practice of reassigning students to new CTA1 sections
\citep{reassignmentEffect}.
In cases such as these, the mediation framework may not apply.

The following subsection reviews the formal definitions of direct and indirect effects.
Subsection \ref{sec:medID} discusses identification, and
\ref{sec:indirectEst} and \ref{sec:directEst} discuss estimation of
indirect and direct effects, respectively.

\subsection{Review: Defining Direct and Indirect Effect}
Express a subject's potential outcomes as a function
of two variables, treatment assignment $Z$ and hint requests $H$:
$Y(Z,H)$.
Now, if $Z$ and $H$ are both binary, as in our example, subjects
each have four potential outcomes:
\begin{equation*}
Y(0,0);\;Y(1,0);\;Y(0,1);\;Y(1,1)
\end{equation*}
representing the outcomes they would exhibit were they assigned to
control and requested few hints, assigned to
treatment and requested few hints, assigned to control and requested
many hints, or assigned to treatment and requested many hints, respectively.
In the CTA1 trial, however, the potential outcome $Y(0,1)$ is
meaningless: students in the control group had no access to CTA1, and
therefore could not request hints.
This will be an important factor going forward.

Since $H$ is itself affected by treatment assignment, it
too has potential values:
\begin{equation*}
H(0);\;H(1)
\end{equation*}
representing the hints that would be requested under treatment and
control.
In the CTA1 study, $H(0)=0$ for all subjects.

Combining potential values for $Y$ and for $H$ yields the fundamental
building blocks of causal mediation analysis
\citep[e.g.][]{vanderweele2015explanation}.
Let $Y(z,H(z'))$
represent the outcome that a subject would express given treatment
assignment $z$, but the hint behavior that he or she would have
expressed under assignment $z'$.
When $z=z'$, the original two potential outcomes for $Y$ emerge:
$Y(1,H(1))=Y(Z=1)$ is the outcome exhibited under the treatment, when $H$
takes its potential treatment value; similarly, $Y(0,H(0))=Y(Z=0)$. On
the other hand, $Y(1,H(0))$ and $Y(0,H(1))$ are strictly
counterfactual. $Y(1,H(0))$ gives the outcome value
that would result for a subject assigned to the treatment condition
but who nevertheless requested hints as he or she would have under
control. $Y(0,H(1))$ gives the potential outcome for a subject
assigned to control but who nevertheless requested hints as he or she
would have under treatment.
This last quantity is problematic in our case, since $H(1)$ can take the value 1,
and as we have seen $Y(0,1)$ is undefined.

This framework facilitates precise definitions of direct and indirect
effects; in fact, there are two versions of each \cite[e.g.][]{imai2011unpacking}:
\begin{align*}
\delta(1)&=Y(1,H(1))-Y(1,H(0))\\
\delta(0)&=Y(0,H(1))-Y(0,H(0))\\
\end{align*}
are the indirect effects, contrasting potential outcomes when $H$
varies as it would with varying treatment assignment, but holding the
assignment itself constant at either 1 or 0.
\begin{align*}
\xi(1)&=Y(1,H(1))-Y(0,H(1))\\
\xi(0)&=Y(1,H(0))-Y(0,H(0))\\
\end{align*}
represent the direct effects: holding the value of $H$ constant,
either at its potential value under treatment or control, while
varying treatment assignment.
The total treatment effect, $Y(1,H(1))-Y(0,H(0))$, can then be
decomposed in two ways: as $\delta(1)+\xi(0)$ or as
$\delta(0)+\xi(1)$.
These decompositions can potentially reveal the role hints play in the
CTA1 treatment effect: $\delta(1)$ gives the extent to which
assignment to CTA1 affects a student's posttest score by (possibly)
causing him or her to request many hints.
$\xi(0)$ gives the effect of assignment to CTA1 if hints were,
perhaps counterfactually, held at their control level, $H=0$.

Rather than attempt to estimate individual direct and indirect effects
$\xi_i$ and $\delta_i$, our goal will be to estimate their means,
$\EE \xi$ and $\EE \delta$.

\subsection{Identification of Direct and Indirect
  Effects}\label{sec:medID}

Since the potential outcome $Y(0,1)$ is undefined in the CTA1 hints
study, any expression (potentially) including $Y(0,1)$ is also
undefined; this includes $\delta(0)$ and $\xi(1)$.
Indeed, since hint behavior is unobserved in the control group
altogether, conventional approaches to mediation analysis do not
apply.
Further, were we to operationalize hint behavior as a continuous
variable, as may seem natural, mediation analysis in this case would
also be nearly impossible.
This is because students assigned to the control group requested exactly
zero hints, whereas merely
5
members of the treatment group
requested no hints over the course of the study, and these students
barely used the tutor at all.
That being the case, the data provide little to no information about
the distribution of potential outcomes when the software is used
but no hints are requested.

Dichotomizing hint requests into $H$ provides a
way forward.
Since the experimental design ensures that $H(0)=0$, we have that $Y(1,H(0))\equiv
Y(1,0)$; this potential outcome is observed for subjects in
the treatment arm with $H=0$.
As above, $Y(0,H(0))=Y(0,0)$ and $Y(1,H(1))$ are the observed outcomes
for subjects in the control and treatment arms, respectively.
Therefore, we observe each of the potential outcomes in the definitions of
$\delta(1)$ and $\xi(0)$ for at least some non-trivial
portion of subjects in the study sample.

Of course, $H=0$ means something different in the two treatment arms:
in the treatment group $H=0$ typically means requesting
\emph{some} hints, but not many, whereas subjects in the control group
request exactly zero hints.
One way to make sense of mediational estimands involving $H$ is to
imagine an experiment in which students are assigned to different
values of $H$.
70\% of students in the treatment group
are randomly assigned to $H=1$ and 30\% to
$H=0$.
This assignment constrains $\bar{h}$, the proportion of problems on which they
may request hints, so that if $H_i=0$,
$\bar{h}_i$ is constrained to be less than
0.6,
but if $H_i=1$ $\bar{h}_i$ must be greater than or equal to
0.6.
In contrast, all students in the control group are assigned $H=0$ and $\bar{h}=0$.

Average potential outcomes $\EE Y(1,H(1))$ and $\EE Y(0,H(0))$ are
identified due to the randomization of $Z$, but estimating $\EE
Y(1,H(0))$ requires imputing the
potential outcome $Y(1,0)$ for treated subjects with $H_i=1$.
However, it turns out that the matching estimators from Section
\ref{sec:observational} already solved this problem:
under matched ignorability, within matched sets, $Y(1,0)$
is independent of $H$, implying that $\EE[Y(1,0)|M,H=1]=\EE[Y(1,0)|M,H=0]$.
That suggests imputations
\begin{equation}\label{eq:estY10}
\hat{Y}_i(1,0)=\begin{cases}
 Y_i & H_i=0\\
\bar{Y}_{M=m_i,H=0} & H_i=1
\end{cases}
\end{equation}
That is, when $Z_i=1$ and $H_i=0$, no imputation is necessary, since
$Y_i(1,0)=Y_i$, the observed outcome. When $Z_i=1$ and $H_i=1$,
impute the average of the observed outcomes from the $H=0$ subjects
matched to $i$ for $Y_i(1,0)$.
Matched ignorability implies that the imputations will be unbiased.

\subsection{Estimating Log Data Indirect Effects}\label{sec:indirectEst}

Since $\EE \delta(1)$ involves only $Z=1$ potential outcomes, and
$H(0)=0$ is known based on the study design, $\EE\delta(1)$ may be
estimated using data from the treatment group only.

Using the imputations from \eqref{eq:estY10}, we estimate
$\EE\delta(1)$ as
\begin{equation*}
\widehat{\EE\delta(1)}=\widehat{\EE Y(Z=1)}-\widehat{\EE
  Y(1,0)}=\frac{1}{n_T}\displaystyle\sum_{i:Z_i=1} Y_i-\hat{Y}_i(1,0)
\end{equation*}
where $n_T=\sum_i Z_i$, the number of subjects in the treatment
group.
Since $Y_i=\hat{Y}_i(1,0)$ whenever $H_i=0$, the summand is non-zero
only when $H_i=1$.
Further, note that when multiple $H=1$ subjects share a
matched set, they also share imputations $\hat{Y}(1,0)$.
Re-writing $\widehat{\EE\delta(1)}$ in terms of matched sets $M=m$,
\begin{align}
\widehat{\EE\delta(1)}&=\frac{1}{n_T}\displaystyle\sum_m\displaystyle\sum_{i:H_i=1,M_i=m}
                        Y_i-\bar{Y}_{M=m_i,H=0} \nonumber \\
&=\frac{1}{n_T}\displaystyle\sum_m
  n_{1m}(\bar{Y}_{M=m_i,H=1}-\bar{Y}_{m=m_i,H=0}) \nonumber \\
&=\frac{1}{n_T}\displaystyle\sum_mn_{1m}\hat{\tau}_m=\frac{\sum_iH_i}{n_T}\widehat{TOT} \label{eq:indTot1}
\end{align}
where $n_{1m}$ and $\hat{\tau}_m$ were defined in Section
\ref{sec:obsEst} as the number of $H=1$ subjects in match $m$ and the
difference in the sample mean of $Y$ between $H=1$ and $H=0$ subjects
in match $m$, respectively, and $\widehat{TOT}$ is the TOT estimate,
\eqref{eq:tauWeightedMean} with $w_m=n_{1m}$.
The upshot of \eqref{eq:indTot1} is that under matched ignorability,
$\EE\delta(0)$ can be estimated as the TOT times the proportion of the
treatment group with $H=1$.

Actually, this relationship between $\EE\delta(0)$ and the TOT holds
in the population, regardless of how the TOT is estimated.
This is because when $H_i=0$, then $H_i(1)=H_i(0)=0$, so the individual direct effect
$\delta_i(1)=0$, and when $H_i=1$,
 $\delta_i(1)=Y_i(H=1)-Y_i(H=0)$, the effect of $H$ on $Y$.
Therefore,
\begin{equation*}
\EE \delta(1)=\EE\EE[\delta(1)|H]=\EE[\delta(1)|H=1]Pr(H=1)=TOTPr(H=1)
\end{equation*}

In the hint study, the definition of $H$ implies that
$Pr(H=1)=\sum_iH_i/n_T=$0.3.
Therefore, our estimate of the average indirect effect is
$\widehat{\EE \delta(1)}=$0.3$\widehat{TOT}$.
In Table \ref{tab:matchResults}, we estimated the TOT as
-0.14 with a
standard error of 0.06.
That implies an average indirect effect of -0.04, with
a standard error of 0.02.

It is worth repeating that these estimates assume matched ignorability---no confounding between
$H$ and $Y$ within matched sets.
With this important caveat, this analysis suggests that the
wide availability of a hints
actually \emph{lowers} the treatment effect, that CTA1 works not
because of hints, but despite them.

\subsection{Estimating Log Data Direct Effects}\label{sec:directEst}

What would the effect be were all students to request few hints?
To answer that question, we estimate average direct effects,
$\EE[\xi(0)]=\EE[Y(1,H(0))-Y(0,H(0))]=\EE[Y(1,0)-Y(0,0)]$.\footnote{This
  is equivalent to the ``controlled direct effect,'' $CDE(0)$
  e.g. \citet[][p. 57]{vanderweele2015explanation}.}
In principal, this could be estimated as the difference in sample
means of $\hat{Y}(1,0)$ in the treatment group and observed $Y=Y(0,0)$
in the control group.
In general, if we let
\begin{equation*}
\tilde{Y}\equiv Z\hat{Y}(1,0)+(1-Z)Y(0,0)
\end{equation*}
then we can estimate $\EE[\xi(0)]$ as the ATE of treatment assignment
$Z$ on $\tilde{Y}$.
Just like estimates of the ATE of $Z$ on $Y$, this estimate should
account for the experimental design: a paired, group-randomized trial
with students clustered in schools and schools randomized within
pairs.

For the sake of simplicity, we will use an OLS regression estimator,
with cluster-robust standard errors \citep{pustejovsky2018small},
clustered at the school level.
(Since in Section \ref{sec:observational} subjects were matched
within schools, school-level cluster-robust standard errors also account for the
dependence of $\hat{Y}(1,0)$ between matched treated students.)
Specifically, we regress:
\begin{equation}\label{eq:directEffReg}
\tilde{Y}_i=\alpha_{p[i]}+\widehat{\EE[\xi(0)]}Z_i+\epsilon_i
\end{equation}
where $\alpha_{p[i]}$ is a fixed intercept for the randomization pair
of subject $i$'s school and $\epsilon_i$ is a regression error.
Using the \texttt{clubSandwich} package in \texttt{R}
\citep{clubsandwich} to estimate standard errors, and 95\% confidence intervals, we estimated a direct
effect of
0.14$\pm$0.25.
Adding covariates to the regression \eqref{eq:directEffReg} estimates a direct effect of
0.18$\pm$0.21.

Table \ref{tab:mediation} summarizes our mediation results: assuming matched ignorability, assignment to CTA1 increases student test scores despite the availability of hints (an indirect effect of -0.04 standard deviations), and the effect due to CTA1's other mechanisms is 0.18.

\begin{table}
\centering
\begin{tabular}{lll}
\hline
Effect&Estimate&CI\\
\hline
Avg. Indirect Effect ($\EE \delta(1)$)& -0.04 & [-0.07,-0.01] \\
Avg. Direct Effect ($\EE\xi$)& 0.14 & [-0.10,0.39] \\
Avg. Direct Effect, covariate adjusted ($\EE\xi$) & 0.18 & [-0.03,0.39] \\

\hline
\end{tabular}
\caption{Estimated effects in mediation analysis. }
\label{tab:mediation}
\end{table}

\section{Principal Stratification}\label{sec:principalStratification}

In a principal stratification \citep[PS;][]{frangakis}  analysis, the researcher stratifies all students in
the RCT---in both the treatment and control arms---based on how they
\emph{would} use the software were they (perhaps counterfactually) assigned
to the treatment condition.
The goal is to estimate the average effect of
assignment to the treatment condition separately in each stratum.
In our analysis, we imagine that each student has his or her own proclivity to request hints, when possible: if assigned to treatment, each student would tend to request hints at a different rate.
How does the effect of assignment to the CTA1 condition vary with these proclivities?
Would the benefit of being assigned to the CTA1 condition be higher or lower for students who would tend to request more hints than for those who would tend to request fewer?

In an RCT, identification of these varying effects relies on
randomization of treatment assignment,
rather than on assumptions about
unmeasured confounders.
Unlike the observational study or mediation analysis, PS in an RCT requires no assumptions about unmeasured confounding.
PS analysis requires data on outcomes
(e.g. posttest scores) in both treatment arms, as well as data on
implementation.
In classical PS
\citep[e.g.][]{page2012principal,feller2016compared}, the intermediate
variable, implementation, is discrete or categorical, leading to
discrete strata.
Other approaches \citep[e.g.]{gilbertHudgens,jin2008principal} allow
the intermediate variable to be continuous.
The method we use here is based on \citet{aoas}, in which the variable
defining principal strata is latent---specifically, the ``ability''
parameter in \eqref{eq:rasch}.
For this approach, implementation is measured with a series of binary
measurements for each student, with each measurement taken from a
specific problem or section.
Other implementation data structures may call for different
measurement models.
Finally, covariates may not be strictly necessary for PS, but they can be extremely helpful.

Below, we describe a Bayesian PS model fit in one step.
However, the full process of PS analysis is a much more involved
procedure.
After a formal introduction in Subsection \ref{sec:psIntro},
Subsection \ref{sec:psModel} describes a set of sub-models for student hint
requests, posttest scores, and treatment effects.
Ideally, each of these models should be carefully tailored to each
data application, and each must be rigorously tested.
Model checking includes examination of residual data plots, fake data
simulation, and fitting a range of alternative specifications in a
sensitivity analysis.
More details for checking PS models can be found in \citet{aoas} and
\citet{psTutorial}, and for Bayesian models in general in
\citet{gelman2014bayesian} and \citet{mcelreath2020statistical}.

\subsection{Principal Effects for EdTech Log Data}\label{sec:psIntro}
In the PS approach, $Y(0)=Y(Z=0)$ and $Y(1)=Y(Z=1)$ are the only potential
outcomes for $Y$.
As in mediation analysis, hint behavior has potential outcomes as
well.
Specifically, the rate at which as student requests hints, $\bar{h}$,
has potential values $\bar{h}(1)$ and $\bar{h}(0)$,
though only $\bar{h}(1)$ is relevant in the CTA1 study.

The goal of PS is to estimate ``principal
effects'':
\begin{equation*}
\tau(r)=\EE[Y(1)-Y(0)|\bar{h}(1)=r]
\end{equation*}
This is the treatment effect for subjects who \emph{would} request
hints at level $\bar{h}(1)=r$, if assigned to the treatment condition.
The goal of PS is to estimate the relationship between $\bar{h}(1)$
and treatment effects; this contrasts with the observational study and mediation
analysis whose goal was the relationship between $\bar{h}$ and posttest scores.
Notably, although $Y(0)$ and $\bar{h}(1)$ and $Y(0)$ are never
simultaneously observed, the $\tau(r)$ estimand does not depend on any
strictly-counterfactual outcomes.
Unlike, say, $Y(0,H(1))$, $Y(0)$ and $Y(1)$ are each truly
potential---they would each occur under treatment assignments $Z=0$
and $Z=1$.
That said, estimating $\tau(r)$ requires estimation of
$\EE[Y(0)|\bar{h}(1)]$, the average of $Y(0)$ over a subset of the
sample that is unobserved and must be inferred.
For this reason, $\tau(r)$ is only \emph{partially} identified
\citep[e.g.][]{mealli2016identification}; even with an infinite
sample, principal effect estimates will still contain uncertainty.

In previous sections $\bar{h}$ was dichotomized into $H$; such
dichotomization is not necessary here.
However, $\bar{h}$ has some disadvantages as a continuous measure of
hint usage \citep{aoas}.
For one, since it is essentially a sample mean over a subset of each student's
worked problems, much of its variance is driven by the total number of
problems students worked---in particular, its extreme values belong to
those students who barely used the tutor at all.
Further, it does not account for varying difficulty of the tutor's
problems.
In Section \ref{sec:dichotomizingHintUsage}, we showed that
dichotomized $\bar{h}$ largely agrees with a dichotomized version of a
more sophisticated measure of hint usage.
However, when modeling hint usage continuously, the correspondence may
not hold \citep[see][for a more complete discussion]{aoas}.

For those reasons, we modeled hint usage at the problem level, with
equation \eqref{eq:rasch}, but with two important differences.
First, a conceptual difference: the $\eta$ in \eqref{eq:rasch} was
replaced by $\eta (1)$, indicating that it is measuring
\emph{potential} hint usage---the hint usage a student would exhibit
were he or she assigned to the treatment condition.
Whereas $\eta$ is only defined for students assigned to the treatment condition,
$\eta(1)$ is defined for all the students in the study.
Second, a modeling difference: instead of the normal mixture model
\eqref{eq:mixture}, we modeled $\eta(1)$ with a normal regression model.
With this change, the PS estimand became
\begin{equation*}
\tau(r)=\EE[Y(1)-Y(0)|\eta(1)=r]
\end{equation*}
the treatment effect for students whose potential hint usage, if assigned to the treatment condition, $\eta(1)=r$.
A full treatment of this variant of PS, including a
discussion of identification and estimation, may be found in
\citet{aoas}.

\subsection{Specifying PS Models}\label{sec:psModel}
Estimating $\tau(r)$ required estimating $\EE[Y(0)|\eta(1)=r]$,
despite the fact that hint requests are never observed at the same time as
$Y(0)$ (in fact $\eta(1)$ isn't directly observed in either the
treatment or control group).
However, among those students with $Z=1$, hint requests are observed
and the conditional distribution $\eta(1)|\bm{x}$ is identified;
because of randomization of $Z$, these inferences extend to the
control group as well \citep{feller2016compared}.

Our approach to PS estimation is model-based and Bayesian, following
\citet{jin2008principal},
\citet{schwartz2011bayesian} and others.
The PS model consists of three sub-models, all depending on a
 vector of parameters $\bm{\theta}$ that contains regression
 coefficients, variance components, and treatment effects.
The sub-models are:
$p(h;\eta(1),\bm{\theta})$, giving the distribution of actual hint
requests as a function of $\eta(1)$, $p(\eta(1)|\bm{x},\bm{\theta})$
giving the distribution of $\eta(1)$ conditional on covariates, and
$p(Y|\eta(1),Z,\bm{x},\bm{\theta})$ giving the conditional
distribution of outcomes.
With these three models in hand, and a prior distribution
$p(\bm{\theta})$, posterior inference for $\bm{\theta}$ proceeds based
on the following structure:
\begin{align*}
p(\bm{\theta}|\bm{Y},\bm{Z},&\bm{X},\{\bm{h}_i\}_{i:Z_i=1})\propto\\
p(\bm{\theta})&\displaystyle\prod_{i: Z_i=1}\int p(Y_i|\bm{x}_i,Z=1,\bm{\theta},\eta(1))p(\bm{h}_{i}|\eta(1),\bm{\theta})p(\eta(1)|\bm{x}_i,\bm{\theta})d\eta(1)\\
\times&\displaystyle\prod_{i:Z_i=0} \int p(Y_i|\bm{x}_i,Z=0,\bm{\theta},\eta(1))p(\eta(1)|\bm{x}_i,\bm{\theta})d\eta(1).\numberthis\label{eq:posteriorEta}
\end{align*}
where $\bm{h}_i$ is the vector of hint request data for subject $i$,
ranging over all challenging problems.
In other words, we estimate parameters by integrating over unknown
(latent) $\eta(1)$ values, using the distribution of $\eta$ estimated
in the treatment group.
This is, in essence, an infinite mixture distribution, with outcome
distribution $p(Y|\cdot)$ and mixing proportions $p(\eta(1)|\cdot)$.
Unlike in typical PS setups, $\eta(1)$ is unobserved for both $Z=1$
and $Z=0$ treatment groups, but is estimated using both $\bm{h}$ and
$\bm{x}$ in the treatment group, but only $\bm{x}$ in the control
group.

First, students' proclivity to request hints is measured by the $\eta$
parameter in (\ref{eq:rasch}).
In the next level, $\eta(1)$ is modeled as a function of baseline
covariates $\bm{x}$:
\begin{equation}\label{eq:rasch2}
\eta(1)|\left(\bm{x}_i,\bm{\theta}\right) \sim
\mathcal{N}\left(\bm{x_i}\bm{\beta}+\epsilon^{tch}_{t[i]}+\epsilon^{scl}_{s[i]},
\sigma\right)
\end{equation}
where $\bm{\beta}$ is a vector of coefficients.
Since students were nested within teachers, who were nested within
schools, we included normally-distributed school ($\epsilon^{scl}$) and
teacher ($\epsilon^{tch}$) random intercepts.
The covariates in the model, $\bm{x}_i$, were detailed in Table
\ref{tab:covariateBalance}; preliminary model checking suggested
including a quadratic term for pretest, which was added as a column of
$\bm{x}_i$.

We modeled students' posttest scores $Y$ as
conditionally normal:
\begin{equation}\label{eq:outcomeSubmodel}
 Y|\left(Z_i,\bm{x}_i,\bm{\theta},\eta(1)\right) \sim  \mathcal{N}\left(
\gamma_{0b[i]}+\bm{x}_i^T\bm{\gamma}+a\eta(1)+Z_i(b_0+b_1\eta(1))+\zeta^{tch}_{t[i]}+\zeta^{scl}_{s[i]},\omega_{Z[i]}\right)
\end{equation}
where $\gamma_{0b[i]}$ is a fixed effect for $i$'s randomization block, $\bm{\gamma}$ are the
covariate coefficients, and $\zeta^{tch}$, and
$\zeta^{scl}$ are normally-distributed teacher and school random
intercepts.
The residual variance $\omega$ varies with treatment assignment $Z$;
this captures measurement error in $Y$, treatment effect heterogeneity
that is not linearly related to
$\eta(1)$, and other between-student variation in $Y$ that is not predicted by
the mean model.

Model \eqref{eq:outcomeSubmodel} implies that treatment effects are
linear in $\eta(1)$,
$\EE[Y(1)-Y(0)|\eta(1)]=\tau({\eta(1)})=b_0+b_1\eta(1)$

While more complex models for $\tau(\eta(1))$ are theoretically
possible (for instance, \citet{jin2008principal} uses a quadratic
model), in our experience non-linear effect models do not perform as
well in model checks as the the linear model.

Covariates $\bm{X}$ were standardized prior to fitting.
Prior distributions for the block fixed effects $\beta^Y_b$ and covariate coefficients
$\bm{\beta^Y}$ and $\bm{\beta^M}$ were normal with mean zero and
standard deviation 2;
priors for treatment effects and the coefficient on $\eta(1)$ were standard
normal.
The rest of the parameters received standard reference priors.
In all cases, we expected true parameter values to be much smaller in
magnitude than the prior standard deviation.

\subsection{Estimating Principal Effects}\label{sec:PSest}

\begin{figure}
\centering
\includegraphics[width=0.95\textwidth]{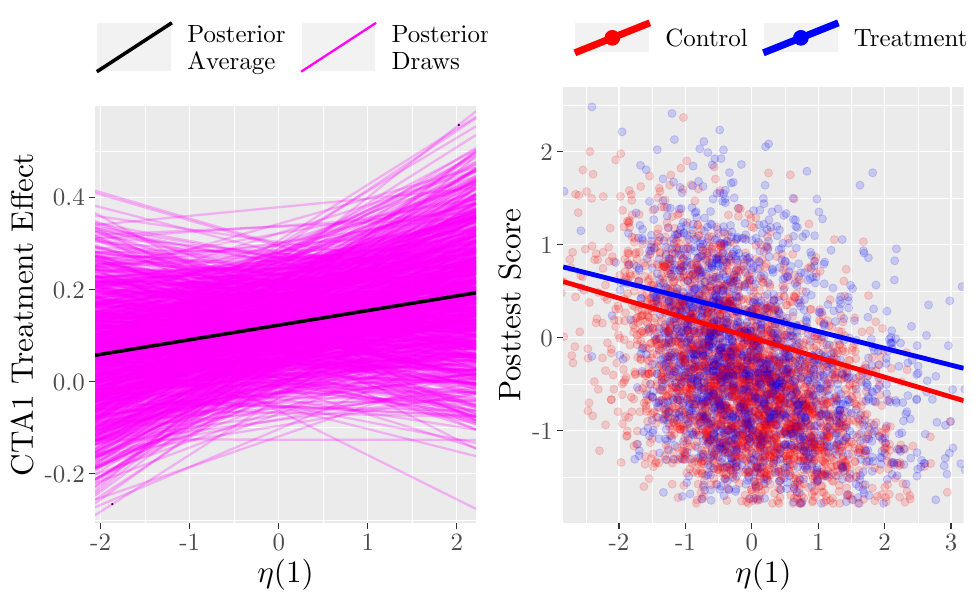}
\caption{Results from the PS model. On the left, the estimated
  CTA1 treatment effect is plotted against hint request proclivity
  $\eta(1)$. The black line is the posterior mean, and the red lines
  are random posterior draws. The right panel plots observed posttest
  scores against one random draw of $\eta(1)$, along with regression
  lines for the treatment and control groups.}
\label{fig:psResults}
\end{figure}

Figure \ref{fig:psResults} shows the estimated linear function
$\hat{\tau}(r)$.
The left panel shows the CTA1 treatment effect, $Y(1)-Y(0)$ as
$r=\tau(1)$ varies; the black line shows the estimate (i.e. posterior
mean) and the magenta lines are random draws from the posterior
distribution, showing estimation uncertainty.
Seventy-five percent of posterior draws of the slope of the
$\hat{\tau}(\cdot)$ function were positive, implying a posterior
probability of 0.75
that students who asked for hints
with greater regularity benefited more from the CTA1 curriculum.
A 95\% highest-density credible interval for the slope parameter is
[-0.06,0.12].
Therefore, the data are consistent with either a slightly negative or
a positive relationship between hint requests and treatment
effects, but the latter is more likely.

The right hand panel plots one posterior draw of $\eta(1)$
alongside observed outcomes $Y$.
The figure also includes estimated regression lines for the two
treatment groups.
Although, as observed in Sections \ref{sec:observational} and
\ref{sec:mediation}, $\eta(1)$ is anticorrelated with $Y$, the slope
is steeper in the control group than in the treatment group.
The distance between the two regression lines is the treatment effect,
which grows with $\eta(1)$.

These results suggest that students who, under treatment, would tend
to ask for hints on challenging problems, are most responsive to
treatment.

\section{Comparing Strategies}\label{sec:synthesis}
In the case discussed here, the observational study and mediation
approaches appear to give the opposite conclusion from principal
stratification---matching and mediation suggest that excessive hint
requests hurt students' posttest scores, while principal
stratification suggests that students who request more hints may
experience higher treatment effects.
This section will compare and contrast the three methods in terms of
 their necessary identification assumptions, their model fitting
 mechanics, and interpretation of their results.

The fundamental difference between the observational study and
mediation approach, on the one hand, and principal stratification on
the other, is in the role of $H$, the intermediate variable.
An observational study or mediation analysis treats $H$ as a causal
agent, an intervention or exposure that affects $Y$.
In mediation analysis, it is also an (intermediate) outcome, affected
by $Z$.
The approach of principal stratification is quite nearly opposite:
$H$ is not an exposure, that ``happens to'' subjects, but rather its
potential value $H(1)$ is a characteristic of the subjects.
Observed $H$ reveals what was there all along.

\subsection{Causal Identification}

Observational studies, and mediation analyses that build on them,
require ignorability assumptions such as Strong Ignorability or
Matched Ignorability.
These may be particularly problematic in our study, since students who
are more likely to request hints are also more likely to struggle with
the material.
We addressed this concern in three ways: first, we focused our attention
on the set of problems in the tutor in which students either requested
a hint or made an error, second, with propensity score matching, and
third with sensitivity analysis.
None of these approaches is bulletproof---they each require further
assumptions about how students request hints.
Eliminating problems in which students showed no signs of struggling
is likely to reduce the association between hint requests and
underlying academic ability, but not eliminate it.
For instance, stronger students may be better able to figure out a
difficult problem without access to hints or error feedback than
weaker students.
Matching depends both on the adequacy of the observed covariates to
capture baseline differences between $H=0$ and $H=1$ students, and the
sensitivity analyses we performed assumed that any un-measured
covariate would predict $H$ and $Y$ no better than pretest or ethnicity indicators, respectively.
These assumptions are impossible to verify from the data at hand;
instead, judging their plausibility requires a keen understanding of student practice within
the tutor.

Mediation analysis builds on the matching design, and requires the same
ignorability assumption; it must
additionally overcome the hurdle of the almost complete lack of
overlap between hint request behavior in the control group (i.e. zero)
and in the treatment group.
Under our approach of dichotomizing hint request behavior, mediation
analysis is essentially a way to interpret the results of the
observational study in the context of the full RCT.

In contrast, no ignorability assumptions are necessary for principal
stratification.
However, principal stratification estimators require modeling
assumptions---most problematically, $p(Y|\bm{x},Z=0,\bm{\theta},\eta(1))$ of
\eqref{eq:posteriorEta}, the model for $Y(0)$
conditional on $\eta(1)$, $\bm{x}$, and parameters $\bm{\theta}$.
While any model specification can, and should, be checked against
observed data, this process can only narrow the space of acceptable
models, it cannot be used to determine the correct form of
$p(Y|\bm{x},Z=0,\bm{\theta},\eta(1))$.
This is because $Y(0)$ and $\eta(1)$ are never observed simultaneously.
Therefore, it is possible for an analyst to misspecify $p(Y|\bm{x},Z=0,\bm{\theta},\eta(1))$, such that principal effect estimates are
severely biased, without being able to detect the model
misspecification with the data.
That is, the data will typically be unable to distinguish between
alternative $p(Y|\bm{x},Z=0,\bm{\theta},\eta(1))$ models that lead to
qualitatively different conclusions.
In practice, we make the untestable assumption that $p(Y|\bm{x},Z=0,\bm{\theta},\eta(1))$ is
of the same form as $p(Y|\bm{x},Z=1,\bm{\theta},\eta(1))$.

An additional contrast of note between matching-based studies and
principal stratification relates to transparency.
However complex the process of devising a match may be, the resulting
design is simple and clear---subjects with $H=1$ are compared against
matches with $H=0$.
This transparency may help researchers assess the plausibility of the
match, and diagnose potential problems with resulting estimates.
Though principal stratification does not depend on ignorability, it does depend on a highly
intricate and parametric modeling structure.
If any part of the model is sufficiently misspecified, the results may
be wrong.
Perhaps worse, \citet{feller2016principal} argued that even well-specified
models can give severely biased estimates.
Extensive model interrogation and checking is necessary for principal stratification
analysis, but the complexity of the principal stratification model makes this process
particularly difficult.
\citet{aoas} gives some examples of potentially fruitful model
checking procedures, which are also carried out in the online supplement.

\subsection{Results and Interpretation}

Our observational study suggested that hint requests have a negative effect: students who
requested hints more often than the 70th
percentile score lower on the posttest, after controlling for observed
covariates.
Taken at face value, this suggests that the optimal strategy is to request few, if any, hints.
However, this conclusion assumes that there were no omitted confounders.
An omitted confounder that is as important as pretest could explain the relationship, but an omitted confounder as important as ethnicity---our second most important covariate---could not (so long as we estimate an OLS-weighted average effect).

Mediation analysis essentially interprets the observational study result in terms of the
overall CTA1 treatment effect.
Specifically, the analysis in \ref{sec:mediation} showed that CTA1 affected students'
posttest scores in (at least) two opposite ways: by allowing hints, it
lowered their posttest scores, but its other mechanisms increased
their scores by an even greater amount.

The
principal stratification analysis also found a negative correlation
between hint requesting and posttest scores.
Those students who request more hints are those who need more help of
some form, and tend to score lower on the posttest.
Fortunately, the principal stratification analysis also suggests (with
probability 0.75) that these students
may benefit more from CTA1 than their peers who request fewer hints.
However, it is unclear whether their larger treatment effects are due
to their higher rate of hint requests, or to some other related characteristic.

Can these three sets of results be reconciled?
Does requesting hints help or hurt?
Technically, principal stratification has nothing to say about the effect of hints---$\eta(1)$
is modeled as a student characteristic, not an intervention.
(See \citealt{jin2008principal} for a nice discussion of this point.)
Still, doesn't the correlation between treatment effects and hint
requests suggest that hints may be beneficial?
Does that contradict the results in Sections \ref{sec:observational}
and \ref{sec:mediation} that hints requests lower students' posttest
scores?
In fact, the results of the three methods can be reconciled.
For instance, it may be that requesting hints indeed lowers test
scores---a negative indirect effect---but that those students who are
likely to request more hints also tend to experience greater direct
effects.
That is, it may be that the observational and mediational results are
correct---that requesting hints lowers test scores---but that the principal stratification
results are also correct---students who request more hints tend to
experience higher treatment effects, due to other mechanisms.
(Of course, the data are also consistent with the possibilities that
the relationship between hint requests and treatment effects is
negative, or close to zero, in which case there is less that needs to
be explained.)

\subsection{Choosing Between Methods}

The previous discussion suggests three criteria for choosing between
the three approaches we've discussed here.
First is the question the researcher seeks to answer.
In their most straightforward interpretations, principal stratification
is a method for examining treatment effect heterogeneity (are students
benefiting differently?), whereas
mediation analysis is a method for assessing causal mechanisms (is the
wide availability of hints in CTA1 beneficial to students?).
That said, principal stratification may also shed light on potential causal mechanisms---if,
indeed, treatment effects are higher for students who request more
hints, then hints may play a role in the tutor's effectiveness.

A second criterion is the researcher's willingness to make
ignorability assumptions.
A researcher's confidence in her understanding of a data generating
process and the quality of observed covariates can translate into
confidence about the results of an observational study or mediation
analysis.
Conversely, researchers unwilling to consider untestable ignorability
assumptions will find principal stratification an attractive alternative.

The final criterion is a researcher's comfort with complex statistical
models.
Researchers who are able and willing to experiment with a range of
models for a dataset, and perhaps devise tests of model fit tailored
to a specific problem, may be more confident in the fit of a principal stratification model.
Researchers who prefer transparency and non-parametric or
semi-parametric analysis may prefer matching studies.

Additional philosophical concerns come in to play as well.
For instance, unlike observational studies and principal stratification, mediation analysis
depends on strictly counterfactual quantities, such as $Y(0,H(1))$,
that can never occur.
Analogous quantities in principal stratification, such as $\EE [Y(0)|\eta(1)]$ are
unobservable, but nevertheless refer to averages of potential outcomes
among actual experimental subjects.
These concerns are important but beyond the scope of the current
paper; for more discussion, see the citations in the introduction.

In summary: there is much to recommend the observational study approach
if risks of confounding bias appear limited
(due to the availability of high-quality covariates, good reason to
believe variation in implementation is random or haphazard, or other
factors).
A wide variety of straightforward and transparent estimation techniques
are available, such as the propensity score
matching illustrated here.
The result---the ATE of implementation on the outcome---is an
intuitive estimand.

When employed, mediation analysis can contextualize the estimated
implementation effect from the observational study in terms of the
experiment's overall treatment
effect, enabling implementation to be interpreted as a causal
mechanism.

However, where unobserved confounding is a serious threat, and complex
statistical modeling (and model checking) is not a practical barrier,
principal stratification may be a better choice because no
ignorability assumptions are necessary.

Of course, there are approaches to modeling implementation beyond
the three considered here. For instance, principal stratification
based on non-parametric bounding \citep{bounding} or randomization
inference \citep{nolen} would avoid the parametric assumptions of
model-based principal stratification, as would a standard moderation
analysis based on predicted implementation, $\hat{\eta}(1)$ instead of
partially observed $\eta(1)$.
We hope that future work will develop these methods,
and still others.



\section{Discussion: Causal Inference and Measurement}\label{sec:conclusion}
This case study has focused on three approaches to causal modeling of
hint requests during an RCT of the Cognitive Tutor program.
We hope we have demonstrated some of the potential and some of the
challenges that new datasets logged by online
learning systems bring to these old problems.

Measurement of students' hint request rates played a central role in all three approaches we considered.
The results were largely driven by three measurement decisions.
First, our decision to consider only worked Cognitive Tutor problems on which students either requested a hint or made an error (or both) reduced the relationship between hint requests and student algebra I ability.
Second, our decision to dichotomize hint request rates in the observational study and mediation analysis allowed us to use of a traditional matching estimator and to identify mediational estimands.
Third, our decision to use a Rasch model to measure hint usage in the principal stratification analysis accounted for differences between students in both the number and difficulty of worked CT problems.

While questions of measurement has always been important in causal inference, analysis of EdTech log data brings them to the fore.
Log datasets from technology products are large, multivariate, and complex, so careful thought is necessary in order to measure implementation constructs of interest.

Even if they are motivated by methodological concerns, decisions about measurement are inherently also decisions about causal questions---estimates using different measurements answer different questions.
Just as researchers must choose a causal approach, such as matching, mediation, or principal stratification, they must choose a measurement approach as well.
As our case study demonstrated, these two decisions are deeply intertwined.

The size, dimension, and complexity of EdTech log data suggests some exciting opportunities for innovative combinations of measurement and causal approaches.
We demonstrated the inclusion of an IRT model in principal stratification; this suggests the possibility of including other measurement models into causal estimators.
Multivariate measurement models that incorporate not only hint requests, but other usage measures such as time spent, actions taken, and errors could yield deep insights about how students use EdTech products and how different usage patterns correspond to different effects.
The development of these causal models will require simultaneous consideration of causal inference and measurement.


\bibliographystyle{plainnat}
\bibliography{ct}

\newpage
\appendix

\section{Rasch Mixture Model of Hint Usage}

To model hint requests with a mixture model, we modeled each hint
request as in \eqref{eq:rasch}.
Then, instead of modeling $\eta_i$ as normal, we assigned it a
two-component normal mixture model:
\begin{equation*}
\eta\sim
p_0\mathcal{N}(\mu_0,\sigma_0)+(1-p_0)\mathcal{N}(\mu_1,\sigma_1)
\end{equation*}
where we constrained $\mu_0<\mu_1$.

\subsection{Stan Code}
\begin{knitrout}
\definecolor{shadecolor}{rgb}{0.969, 0.969, 0.969}\color{fgcolor}\begin{kframe}
\begin{verbatim}
## data{
## //Sample sizes
##  int<lower=1> N; // total number of obs (long format)
##  int<lower=1> nstud; // # students
##  int<lower=1> nsec; // # sections (problems nested in sections)
## 
## // indices
##  int<lower=1,upper=nstud> studentM[N]; //for each obs, which student?
##  int<lower=1,upper=nsec> section[N]; // for each obs, which section?
## 
## // data data
##  int<lower=0,upper=1> hint[N]; // for each obs, was a hint requested?
## 
## }
## parameters{
## 
##  vector[nstud] eta; // ability parameter
##  real<lower=0,upper=1> p1; // mixing proportion
##  //ordered[2] mu;             // locations of mixture components
##  real<upper=0> mu0; //location of mixture component 0
##  vector<lower=0>[2] sigma;  // scales of mixture components
## 
##  real delta[nsec]; // section difficulty parameter
##  //real<lower=0> sigSec; // scale of difficulty parameters
## 
## }
## transformed parameters{
##  real<lower=0> mu1; // location of mixture component 1
##  mu1= -(1-p1)/p1*mu0;
## }
## model{
##  // logs of mixing proportions to save time
##  real log_p1=log(p1);
##  real log_p0=log(1-p1);
## 
## // linear predictor for logit model
##  real linPred[N];
## 
## // priors
##  sigma ~ lognormal(0, 2);
##  mu0 ~ normal(0, 2);
## // sigSec ~ lognormal(0,2);
##  p1~ beta(2,2); // I'd rather the groups have similar sizes
## 
## // model for section difficulty parameters
##  delta~normal(0,3);//sigSec);
## 
## // Rasch model for hint requests
##  for(i in 1:N)
##   linPred[i]= eta[studentM[i]]-delta[section[i]];
## 
##  hint~bernoulli_logit(linPred);
## 
## 
## // mixture model for ability parameters
##  for(n in 1:nstud)
##   target += log_sum_exp(log_p0+normal_lpdf(eta[n]|mu0,sigma[1]),
## 			log_p1+normal_lpdf(eta[n]|mu1,sigma[2]));
## 
## }
\end{verbatim}
\end{kframe}
\end{knitrout}
\section{Covariate Balance Before and After Matching}
\begin{table}[ht]
\centering
\begin{tabular}{lrrrrrr}
  &\multicolumn{3}{c}{Unmatched}&\multicolumn{3}{c}{Matched}\\
 \hline
 & std.diff & z &  & std.diff & z &  \\ 
  \hline
grade9 & -0.34 & -7.05 & *** & 0.03 & 0.44 &     \\ 
  gradehigher & 0.34 & 7.05 & *** & -0.03 & -0.44 &     \\ 
  raceWhiteAsian & -0.64 & -12.83 & *** & 0.03 & 0.75 &     \\ 
  raceBlackMulti & 0.23 & 4.71 & *** & 0.01 & 0.24 &     \\ 
  raceHispAIAN & 0.51 & 10.35 & *** & -0.05 & -0.98 &     \\ 
  pretest & -0.52 & -10.54 & *** & 0.09 & 1.82 & .   \\ 
  spectypical & -0.16 & -3.32 & *** & 0.03 & 0.48 &     \\ 
  specspeced & 0.34 & 7.07 & *** & -0.10 & -1.44 &     \\ 
  specgifted & -0.14 & -3.03 & **  & 0.06 & 1.16 &     \\ 
  esl0 & -0.37 & -7.70 & *** & 0.04 & 0.63 &     \\ 
  esl1 & 0.37 & 7.70 & *** & -0.04 & -0.63 &     \\ 
  stateCT & 0.58 & 11.81 & *** & -0.00 & 0.00 &     \\ 
  stateKY & -0.36 & -7.51 & *** & -0.00 & 0.00 &     \\ 
  stateLA & -0.08 & -1.68 & .   & -0.00 & 0.00 &     \\ 
  stateMI & -0.36 & -7.51 & *** & 0.00 & 0.00 &     \\ 
  stateTX & 0.43 & 8.77 & *** & 0.00 & 0.00 &     \\ 
   \hline
\end{tabular}
\caption{Covariate balance (standardized differences and Z-scores) before and after the propensity score match.} 
\label{tab:balance}
\end{table}

\begin{figure}[!h]
\centering
\includegraphics[width=0.5\textwidth]{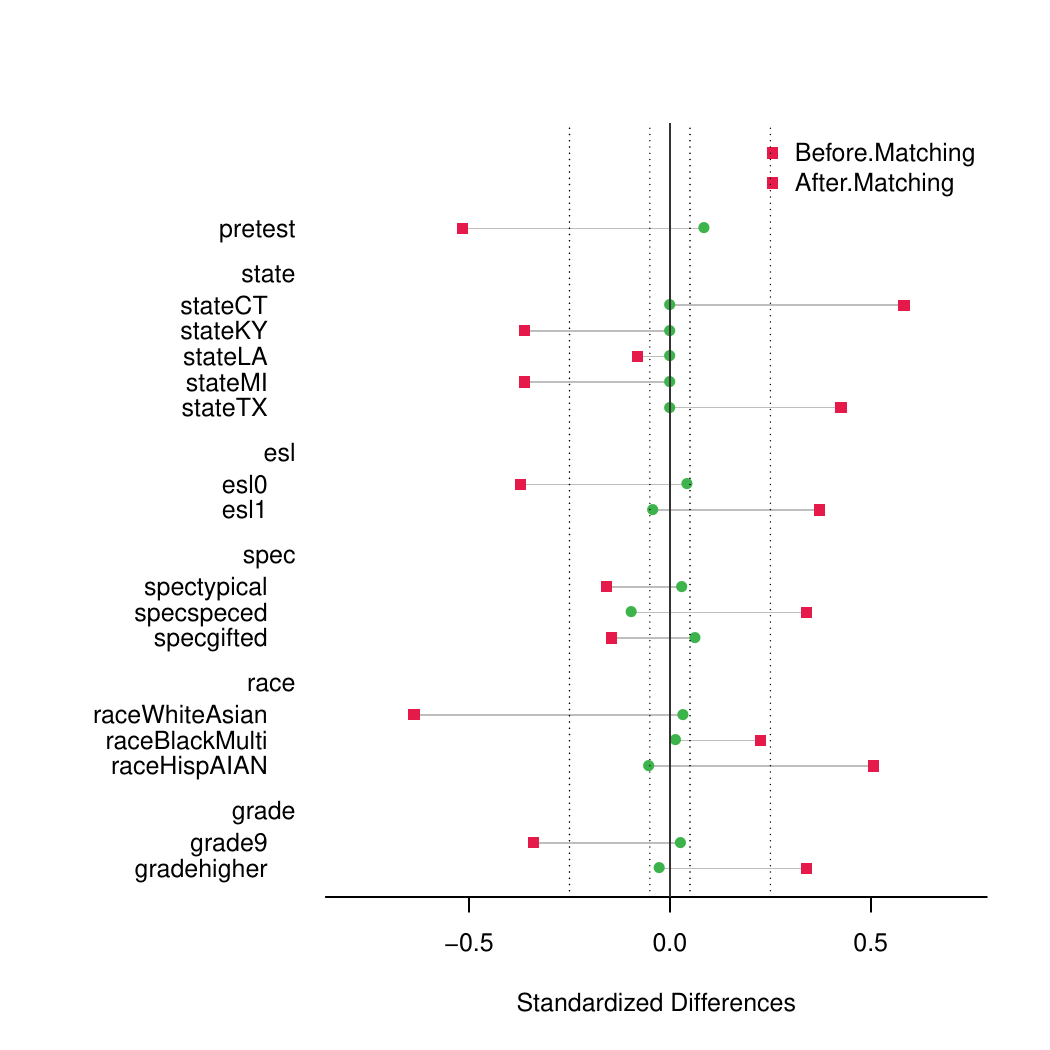}
\caption{Covariate balance (standardized differences) before and after
  the propensity score match.}
\label{fig:balance}
\end{figure}

\section{Stan Code for Principal Stratification Model}

\subsection{Stan Code}
\begin{knitrout}
\definecolor{shadecolor}{rgb}{0.969, 0.969, 0.969}\color{fgcolor}\begin{kframe}
\begin{verbatim}
## data{
## //Sample sizes
##  int<lower=1> nsecWorked;
##  int<lower=1> ncov;
##  int<lower=1> nstud;
##  int<lower=1> nteacher;
##  int<lower=1> nsec;
##  int<lower=1> nschool;
##  int<lower=1> npair;
## 
## // indices
##  int<lower=1,upper=nteacher> teacher[nstud];
##  int<lower=1,upper=npair> pair[nstud];
##  int<lower=1,upper=nschool> school[nstud];
##  int<lower=1,upper=nstud> studentM[nsecWorked];
##  int<lower=1,upper=nsec> section[nsecWorked];
## 
## // data data
##  int<lower=0,upper=1> hint[nsecWorked];
##  matrix[nstud,ncov] X;
##  int<lower=0,upper=1> Z[nstud];
##  real Y[nstud];
## 
## }
## parameters{
## 
##  vector[nstud] eta;
## 
##  vector[ncov] betaU;
##  vector[ncov] betaY;
## 
##  real a1;
##  real b0;
##  real b1;
## 
##  real teacherEffY[nteacher];
##  real teacherEffU[nteacher];
##  real pairEffect[npair];
##  real schoolEffU[nschool];
##  real schoolEffY[nschool];
##  real delta[nsec];
## 
##  real<lower=0> sigTchY;
##  real<lower=0> sigSclY;
##  real<lower=0> sigY[2];
##  real<lower=0> sigTchU;
##  real<lower=0> sigSclU;
##  real<lower=0> sigU;
## }
## 
## model{
##  real linPred[nsecWorked];
##  vector[nstud] muY;
##  vector[nstud] muU;
##  real useEff[nstud];
##  real trtEff[nstud];
##  real sigYI[nstud];
## 
## 
## // hint model
##  for(i in 1:nsecWorked)
##   linPred[i]= delta[section[i]]+eta[studentM[i]];
## 
##  for(i in 1:nstud){
##   useEff[i]=a1*eta[i];
##   trtEff[i]=b0+b1*eta[i];
##   muU[i]=teacherEffU[teacher[i]]+schoolEffU[school[i]];
##   muY[i]=teacherEffY[teacher[i]]+schoolEffY[school[i]]+
##          pairEffect[pair[i]]+useEff[i]+Z[i]*trtEff[i];
##   sigYI[i]=sigY[Z[i]+1];
##  }
## 
##  //priors
##  betaY~normal(0,2);
##  betaU~normal(0,2);
##  pairEffect~normal(0,2);
## 
##  a1~normal(0,1);
##  b0~normal(0,1);
##  b1~normal(0,1);
## 
## 
##  schoolEffY~normal(0,sigSclY);
##  schoolEffU~normal(0,sigSclU);
##  teacherEffU~normal(0,sigTchU);
##  teacherEffY~normal(0,sigTchY);
## 
##  hint~bernoulli_logit(linPred);
## 
##  eta~normal(muU+X*betaU,sigU);
##  Y~normal(muY+X*betaY,sigYI);
## }
## generated quantities{
## // int<lower=0,upper=1> hintRep[nsecWorked];
##  real Yrep[nstud];
## 
## // hintRep=bernoulli_logit_rng(linPred);
##  for(i in 1:nstud)
##   Yrep[i] = normal_rng(teacherEffY[teacher[i]]+schoolEffY[school[i]]+
##             pairEffect[pair[i]]+a1*eta[i]+Z[i]*(b0+b1*eta[i])+
## 	    X[i,]*betaY,sigY[Z[i]+1]);
## }
\end{verbatim}
\end{kframe}
\end{knitrout}

\section{Diagnostic Plots for Principal Stratification}

\subsection{Posterior Predictive Density Plots}
\begin{figure}[!h]
\centering
\includegraphics[width=0.5\textwidth]{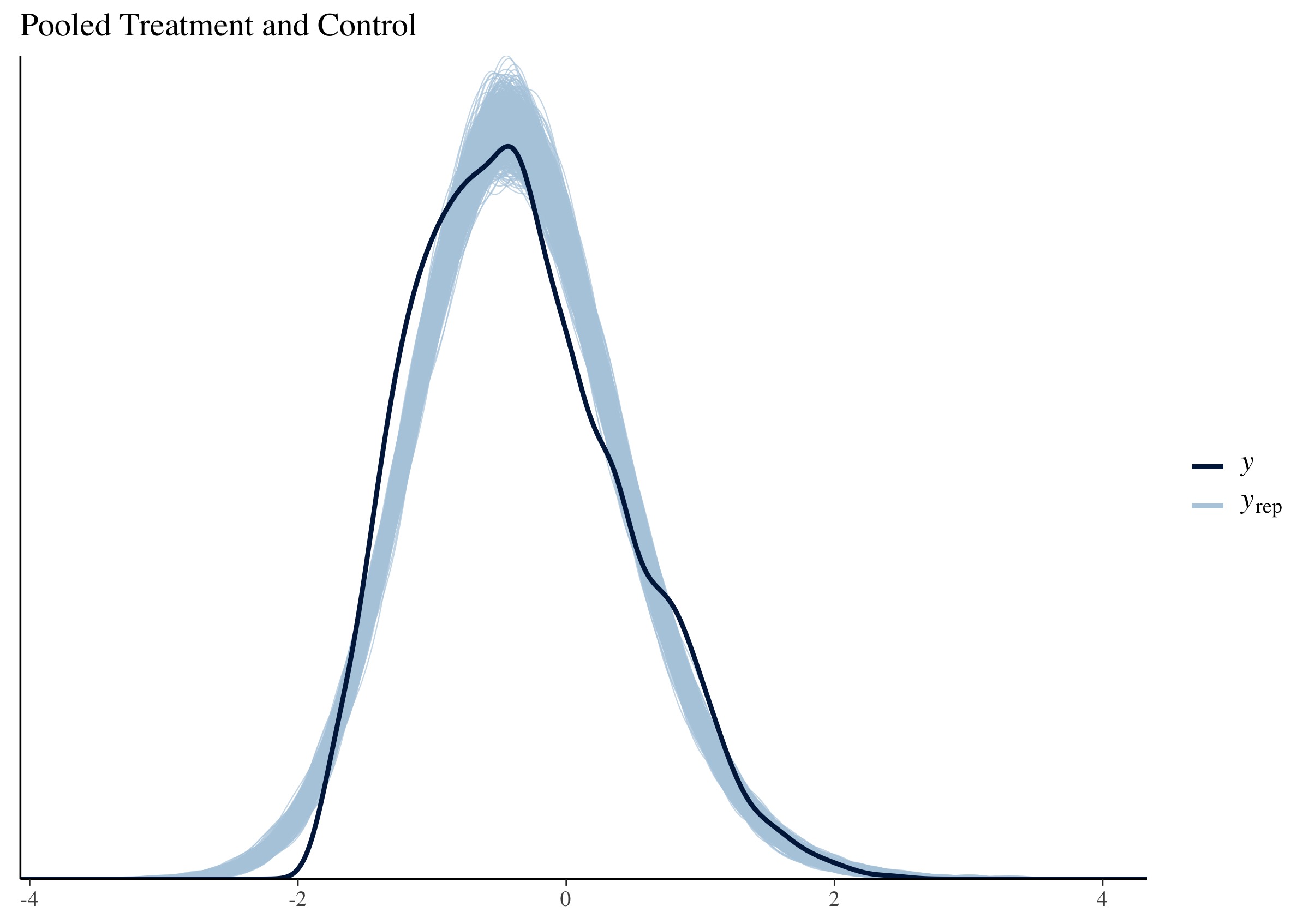}
\caption{A comparison of the estimated density of observed outcomes
  $y$ with 1000 random draws of the model-implied outcome density,
  for the full dataset.}
\label{fig:ppdPooled}
\end{figure}

\begin{figure}[!h]
\centering
\includegraphics[width=0.5\textwidth]{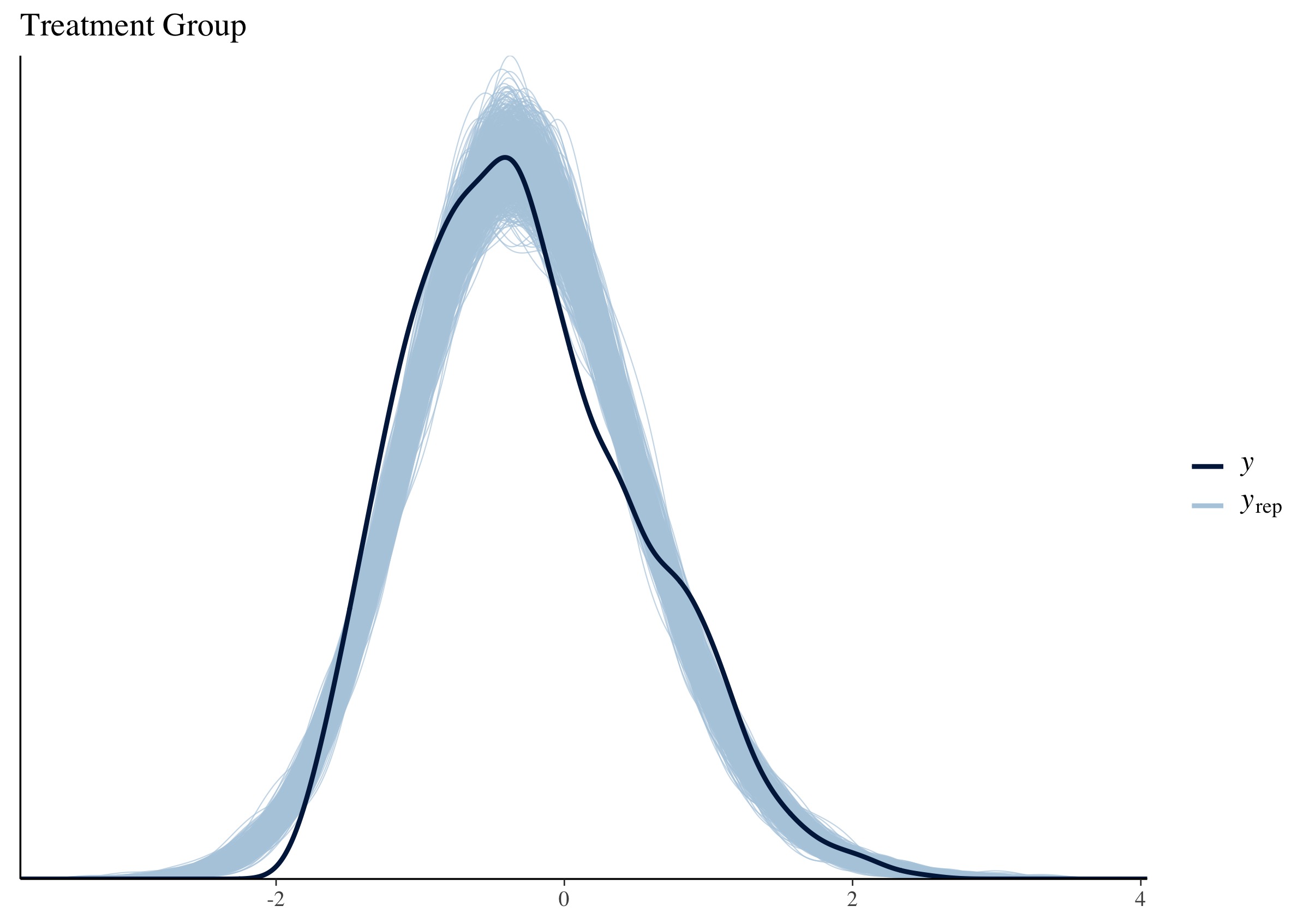}
\caption{A comparison of the estimated density of observed outcomes
  $y$ with 1000 random draws of the model-implied outcome density,
  for the treatment group.}
\label{fig:ppdTrt}
\end{figure}

\begin{figure}[!h]
\centering
\includegraphics[width=0.5\textwidth]{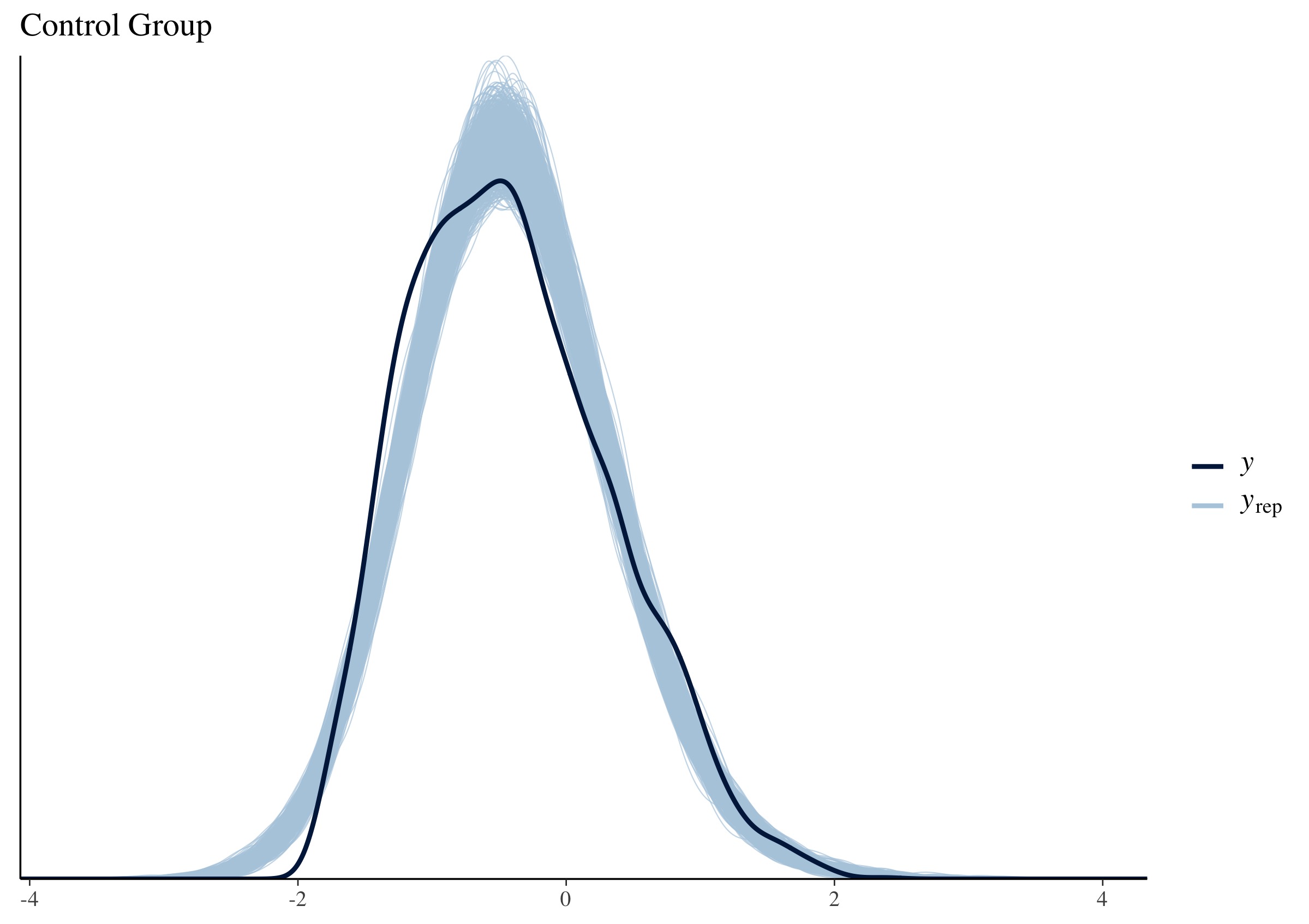}
\caption{A comparison of the estimated density of observed outcomes
  $y$ with 1000 random draws of the model-implied outcome density,
  for the control group.}
\label{fig:ppdCtl}
\end{figure}

Figures \ref{fig:ppdPooled}, \ref{fig:ppdTrt}, and \ref{fig:ppdCtl}
compare the density of observed outcomes $y$ to 1000 random draws of
the model-implied density, $y_{rep}$ for the full (``pooled'') sample,
the treatment group, and the control group, respectively.
They show a close, but not perfect, correspondence between the
observed data and the model.
The most noticeable difference is the left-truncation of the observed
outcomes---the lower tail is lighter than would be expected from a
normal distribution---suggesting floor effects on the posttest not
considered in the model.
The misspecification appears similar in both treatment groups.
A model accounting for floor effects may fit the data more closely.
These plots were made with the \texttt{bayesplot} package
\citep{bayesplot} in \texttt{R}.

\subsection{Residual Plots}
\begin{figure}[!h]
\centering
\includegraphics[width=\textwidth]{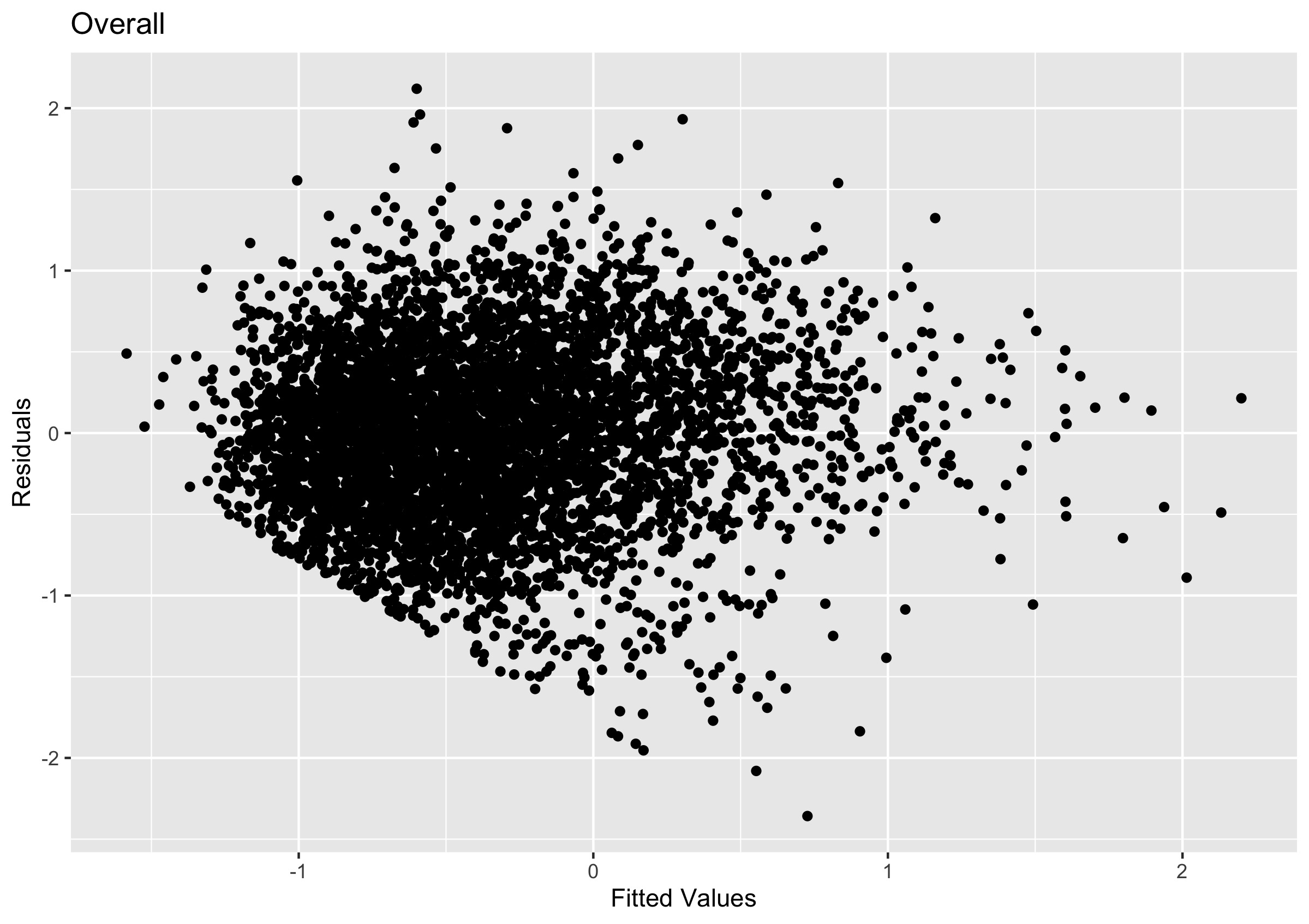}
\caption{A residual plot for the full sample.}
\label{fig:residFull}
\end{figure}

\begin{figure}[!h]
\centering
\includegraphics[width=\textwidth]{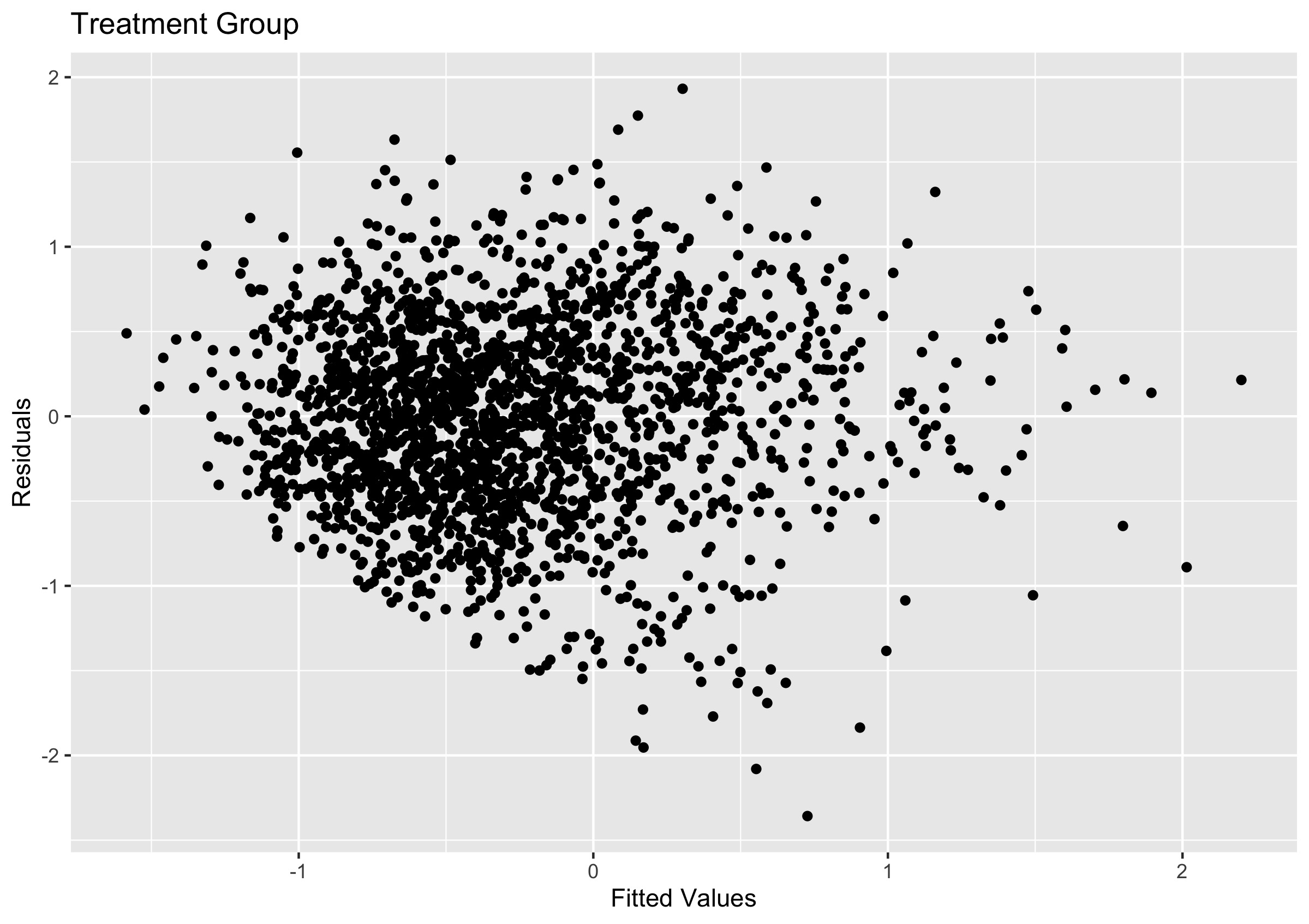}
\caption{A residual plot for the treatment group.}
\label{fig:residTrt}
\end{figure}

\begin{figure}[!h]
\centering
\includegraphics[width=\textwidth]{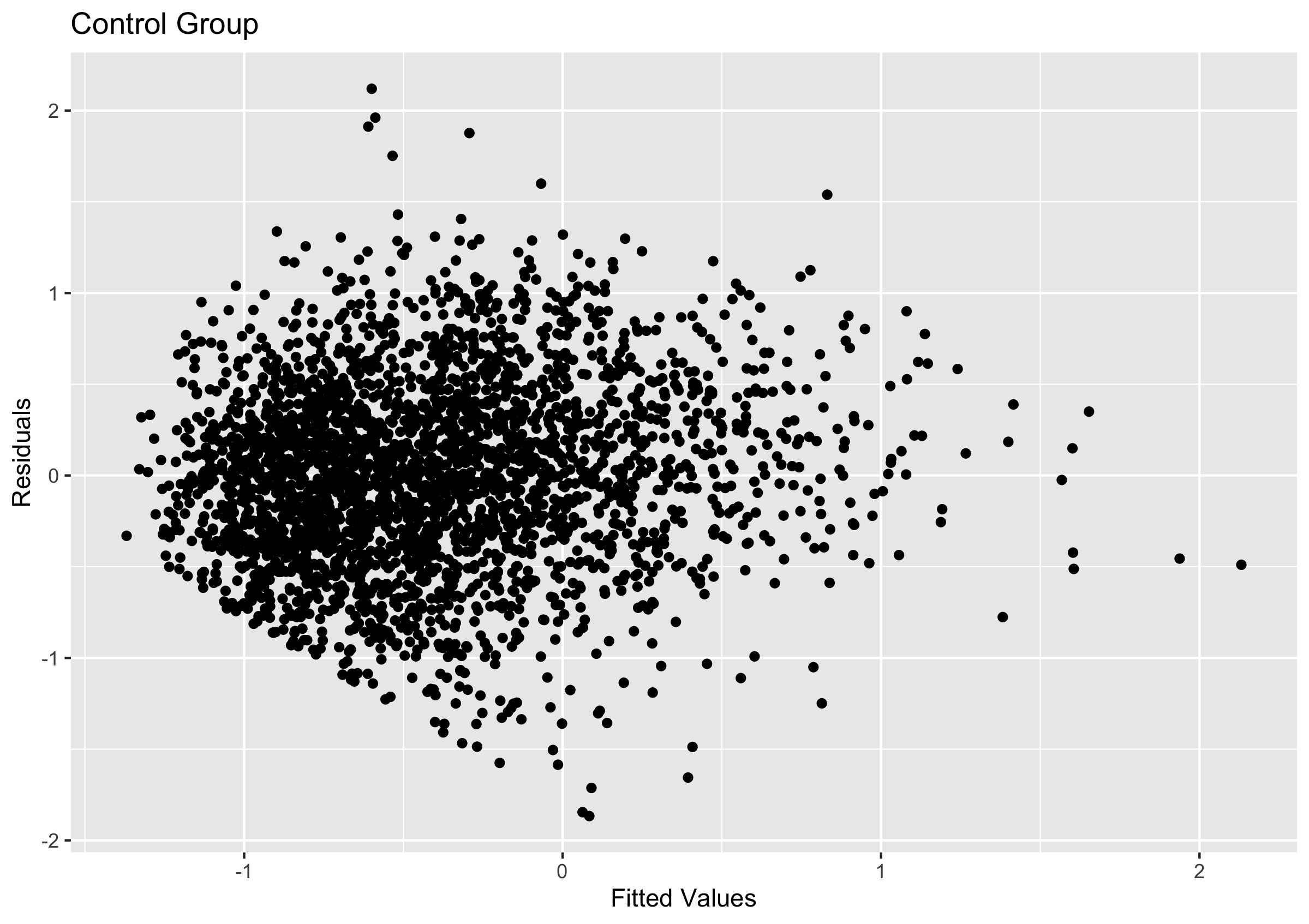}
\caption{A residual plot for the control group}
\label{fig:residCtl}
\end{figure}

Figures \ref{fig:residFull}, \ref{fig:residTrt}, and
\ref{fig:residCtl} plot residuals $Y_i-\hat{Y}_i$ against fitted
values $\hat{Y}_i$ for the full sample, the treatment group, and the
control group, respectively.
Here, fitted values $\hat{Y}$ are posterior means---averages over all
of the draws from the model.
The residual plots reflect the floor effects apparent in the posterior
predictive density plots---a subset of points appears to fall along a
straight line in the bottom left of the plot.
Otherwise, little structure is apparent.

\begin{figure}[!h]
\centering
\includegraphics[width=\textwidth]{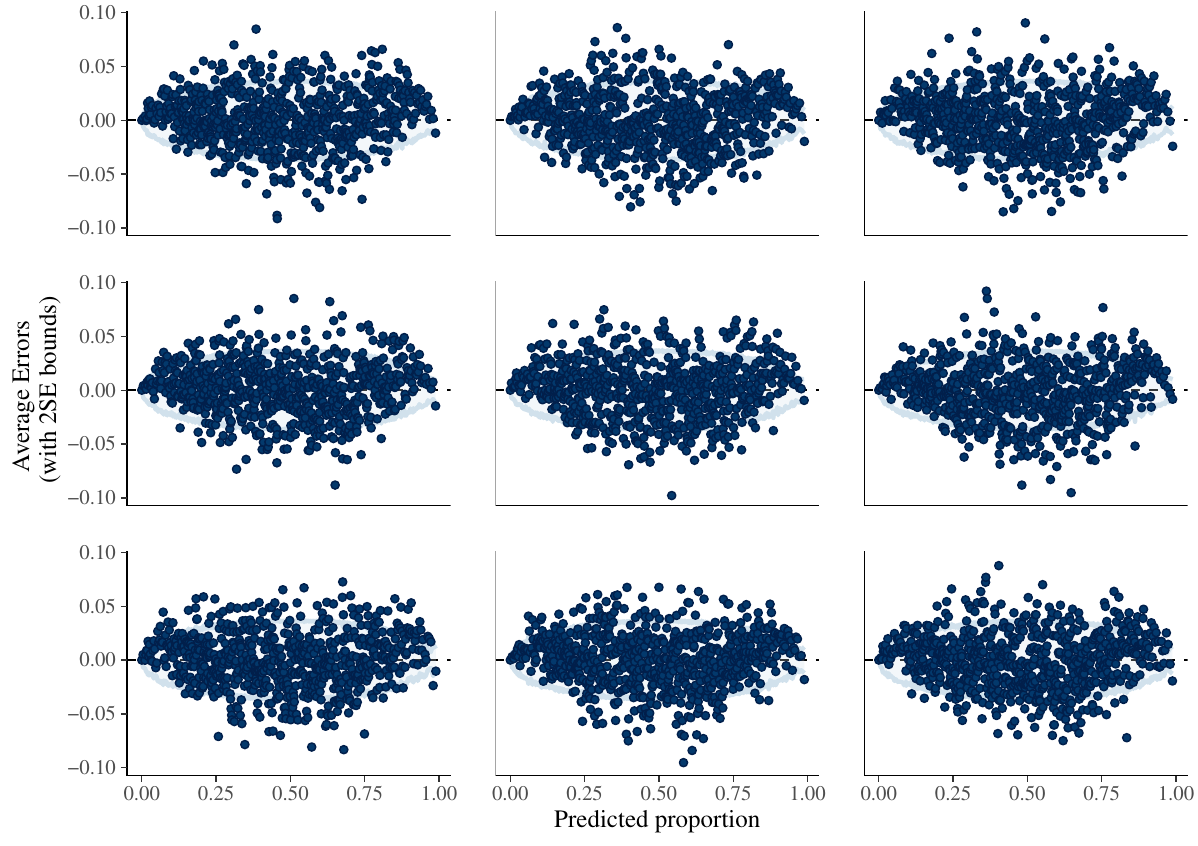}
\caption{Binned residual plot for model of hint requests.}
\label{fig:binnedResids}
\end{figure}

Figure \ref{fig:binnedResids} assesses the hint sub-model,
\eqref{eq:rasch} and \eqref{eq:rasch2} by comparing the binned model-implied
probabilities of hint requests on individual worked problems to residuals---the
differences between these probabilities and the proportions of worked
problems within each bin that resulted in an
actual hint request \citep[c.f.][]{gelman2014bayesian}.
Nine random draws from the posterior distribution are shown.
Little structure is apparent in the binned residuals, suggesting that
the model for hint requests fits well.

\subsection{Fake Data Simulation}
\begin{figure}[!h]
\centering
\includegraphics[width=\textwidth]{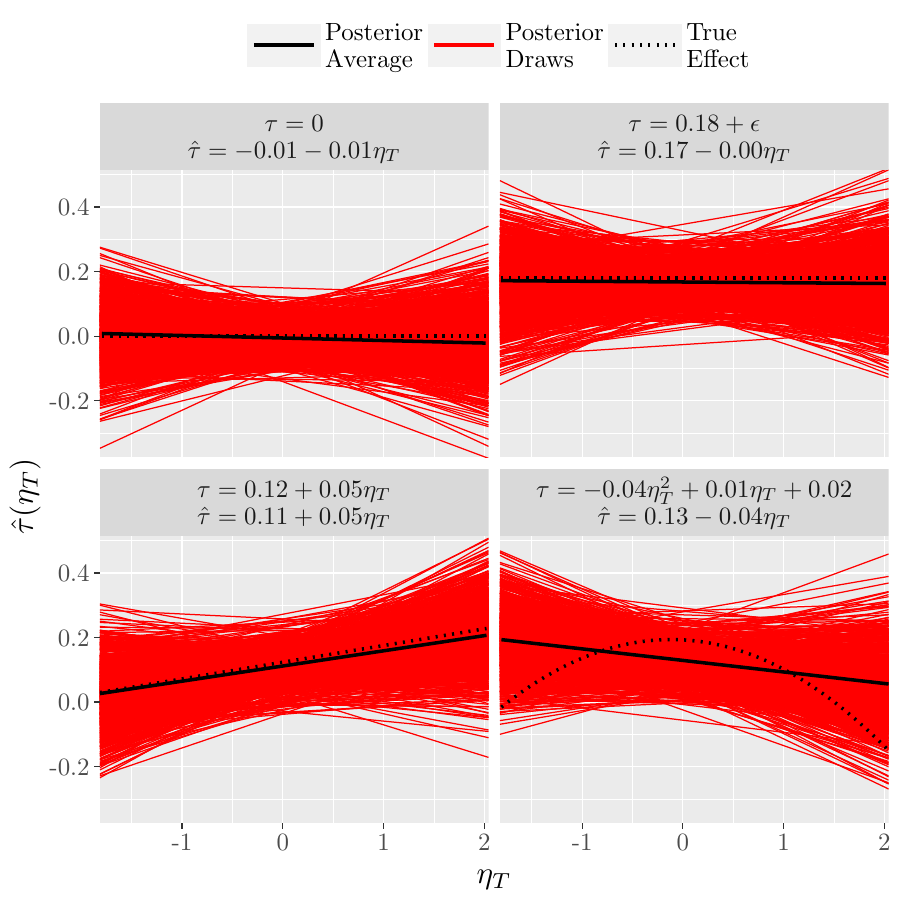}
\caption{Comparisons between principal stratification model estimates
  and true treatment effects when the treatment group is replicated
  (without corresponding hint request data) and relabeled as control,
  and (clockwise from upper left) no treatment effect, a
  random treatment effect uncorrelated with $\eta(1)$, a treatment
  effect linear in $\eta(1)$, or a treatment effect quadratic in
  $\eta(1)$ is added to the original treatment group.}
\label{fig:fakeSim}
\end{figure}

Figure \ref{fig:fakeSim} implements a model assessment for principal
stratification suggested in \citet{aoas}.
In this method, we replace data from the real control group with duplicated
outcome and covariate data from the treatment group (now re-labeled as
control).
In other words, each member of the treatment group appears in the
dataset twice, once in the ``treatment'' group and once in the
``control'' group.
Only the ``treatment'' observations are linked with hint data.
We also re-labeled teachers and schools in the ``control'' group so
that the were distinct from the corresponding labels in the
``treatment'' group.

The upper left plot of \ref{fig:fakeSim} fits the same principal
stratification model described in Section \ref{sec:psModel} to the new
dataset, in which there is exactly no effect.
In the other three panels of \ref{fig:fakeSim}, the model is fit to
the fake data with simulated treatment effects added: a random
treatment effect uncorrelated with $\eta(1)$, a treatment effect
linear in $\eta(1)$ (as we estimated) and a treatment effect quadratic
in $\eta(1)$.
These fake datasets contain the same quirks as the real treatment
group data (and hence the same model misspecification) but with known
treatment effects.
In the first three plots, our model recovered the truth, albeit with
considerable uncertainty.
In the bottom left (quadratic) plot, our model could not recover the
true relationship between $\eta(1)$ and $\tau$ because we modeled
$\tau$ as linear in $\eta(1)$. However, the model did not estimate
a spurious positive or negative associations between $\eta(1)$ and $\tau$.

\end{document}